\documentclass[acmsmall]{acmart}
\usepackage{multirow}
\usepackage{array}

\newcommand{\rev}[1]{\textcolor{black}{\begingroup#1\endgroup}}

\AtBeginDocument{%
  }

\setcopyright{cc}
\setcctype{by}
\acmJournal{PACMHCI}
\acmYear{2026} \acmVolume{10} \acmNumber{2} \acmArticle{CSCW008}
\acmMonth{4} \acmPrice{} \acmDOI{10.1145/3788044}

\usepackage[nolist]{acronym}
\begin{acronym}
    \acro{HCI}{Human-Computer Interaction}
    \acro{CSCW}{Computer Supported Cooperative Work}
    \acro{LLM}{Large Language Model}
    \acro{CS}{Computer Science}
    \acro{RAG}{Retrieval-Augmented Generation}
    \acro{AI}{Artificial Intelligent}
    \acro{MMR}{Maximal Marginal Relevance}
\end{acronym}
\usepackage{dirtytalk}
\usepackage[draft]{fixme}
\usepackage{soul}

\begin{document}

\author{Junling Wang}
\email{junling.wang@ai.ethz.ch}
\orcid{0000-0002-4526-2907}
\affiliation{%
  \institution{ETH Zurich}
  \city{Zurich}
  \country{Switzerland}
}

\author{Lahari Goswami}
\email{lahari.goswami@inf.ethz.ch}
\orcid{0000-0002-8975-5885}
\affiliation{%
  \institution{ETH Zurich}
  \city{Zurich}
  \country{Switzerland}
}

\author{Gustavo Kreia Umbelino}
\email{gus.umbelino@inf.ethz.ch}
\orcid{0000-0001-7754-9606}
\affiliation{%
  \institution{ETH Zurich}
  \city{Zurich}
  \country{Switzerland}
}

\author{Kiara Chau}
\email{kchaugarcia@student.ethz.ch}
\orcid{0009-0005-7966-8787}
\affiliation{%
  \institution{ETH Zurich}
  \city{Zurich}
  \country{Switzerland}
}

\author{Mrinmaya Sachan}
\email{mrinmaya.sachan@inf.ethz.ch}
\orcid{0000-0001-8787-8681}
\affiliation{%
  \institution{ETH Zurich}
  \city{Zurich}
  \country{Switzerland}
}

\author{April Yi Wang}
\email{april.wang@inf.ethz.ch}
\orcid{0000-0001-8724-4662}
\affiliation{%
  \institution{ETH Zurich}
  \city{Zurich}
  \country{Switzerland}
}

\renewcommand{\shortauthors}{Junling Wang et al.}
\newcommand{\AW}[1]{\textcolor{blue}{\textbf{*April*}: #1}}
\newcommand{\JW}[1]{\textcolor{purple}{\textbf{*Junling*}: #1}}
\newcommand{\GU}[1]{\textcolor{red}{\textbf{*Gus*}: #1}}
\newcommand{\MS}[1]{\textcolor{purple}{\textbf{*Mrinmaya*}: #1}}
\newcommand{\LG}[1]{\textcolor{teal}{\textbf{*Lahari*}: #1}}

\newcommand{\inlinequote}[1]{``\textit{#1}''}

\begin{abstract}
LLM-based chatbots like ChatGPT have become popular tools for assisting with coding tasks. 
However, they often produce isolated responses and lack mechanisms for social learning or contextual grounding. 
In contrast, online coding communities like Kaggle offer socially mediated learning environments that foster critical thinking, engagement, and a sense of belonging. 
Yet, growing reliance on LLMs risks diminishing participation in these communities and weakening their collaborative value.
To address this, we propose \concept{}, a design paradigm that embeds social learning dynamics into LLM-based chatbots by surfacing user-generated content and \features{} from online coding communities. 
Using this paradigm, we implemented a RAG-based AI chatbot leveraging resources from Kaggle to validate our design.
Across two empirical studies involving 28 and 12 data science learners, respectively, we found that \concept{} significantly enhances user trust, encourages \rev{engagement} with community, and effectively supports learners in solving data science tasks. 
We conclude by discussing design implications for AI assistance systems that bridge---rather than replace---online coding communities.
\end{abstract}


\newcommand{\sys}{\textsc{ChatCommunity}}
\newcommand{\concept}{Community-Enriched AI}
\newcommand{\feature}{social design feature}
\newcommand{\features}{social design features}
\newcommand{\condone}{RAG(source)}
\newcommand{\condtwo}{RAG(no source)}
\newcommand{\condthree}{GPT-4o}
\newcommand{\condfour}{Browsing}
\newcommand{\commu}{online learning communities}
\title[Bridging Online Coding Communities with AI through Community-Enriched Chatbot Designs]{Bridging Instead of Replacing Online Coding Communities with AI through Community-Enriched Chatbot Designs}




\begin{teaserfigure}
    \centering
    \includegraphics[width=.98\textwidth]{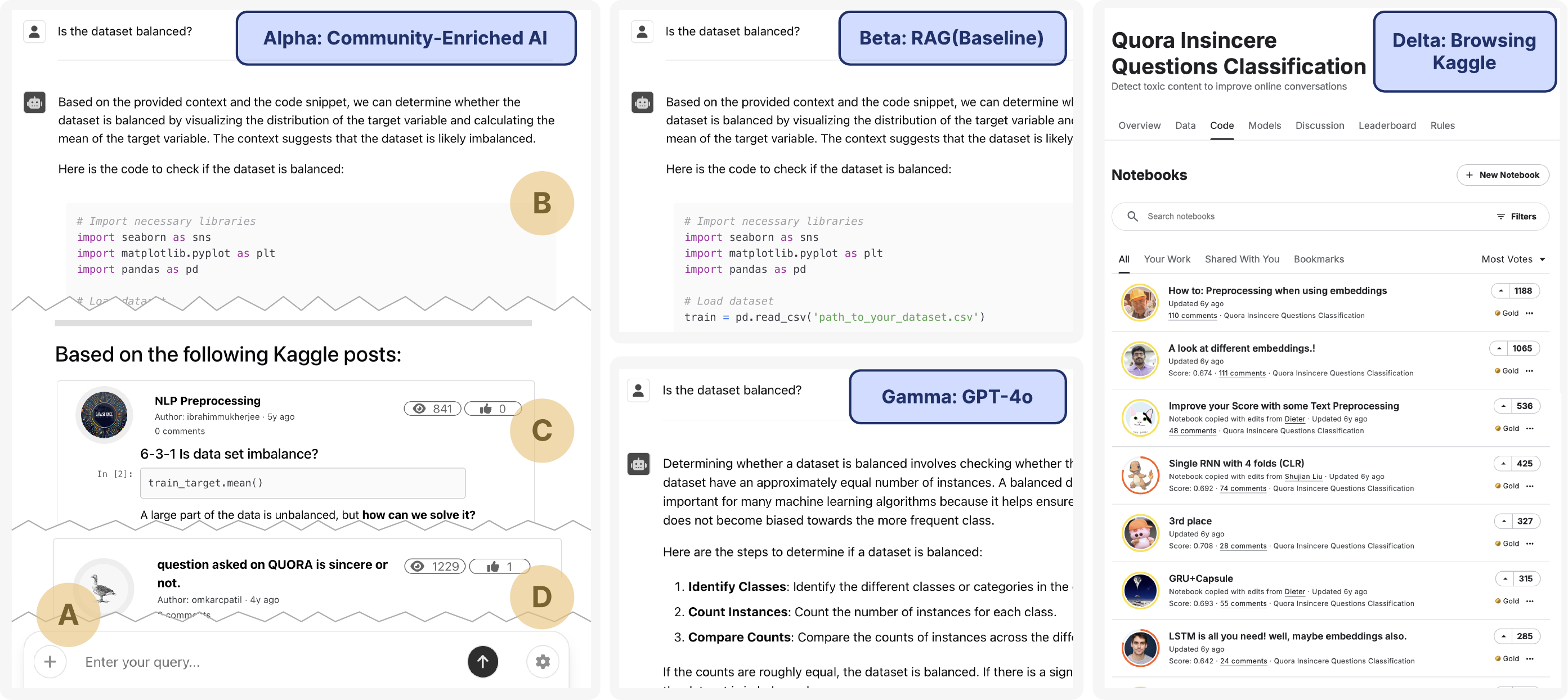}
    \vspace{-10pt}
    \caption{
    An overview of \sys{}, a \concept{} chatbot designed to integrate user-generated content and \features{} from online coding communities into AI-assisted interactions. The interface consists of four key components: (A) a query input bar with session controls, (B) a chat area where AI-generated responses appear, (C) a source document panel previewing the Kaggle posts referenced in responses, complete with \features{} such as author identity and engagement metrics, and (D) an advanced search panel for customizing search parameters. We also illustrate the four conditions used in our evaluation: Alpha (\concept{}), which embeds social learning dynamics into the chatbot interface; Beta (RAG-baseline), which uses the same RAG-based generation model but does not visualize community content; Gamma (GPT-4o), a state-of-the-art LLM baseline without community grounding; and Delta (Browsing Kaggle posts), where users manually explore Kaggle without AI assistance.
    }
    \Description{}
    \label{fig:fig1}
    \vspace{1em}
\end{teaserfigure}

\maketitle
\section{Introduction}
Online coding communities like Kaggle and Stack Overflow are thriving ecosystems where millions of programmers and data scientists engage socially to exchange knowledge, seek solutions, and collaboratively build expertise~\cite{cheng2020, huang2022}.
By 2024, Kaggle alone hosts nearly 19 million active users~\cite{kaggleKaggleUsers}, serving not merely as repositories of knowledge but as social spaces that facilitate knowledge exchanges and co-creation through meaningful social interactions.
Social interactions within these communities offer several distinct benefits for data science learners: they expose learners to diverse problem-solving approaches~\cite{tausczik2017,cheng2020}, surfacing active discussions and debates that promote critical thinking and reflection~\cite{tanprasert2024_cscw,walker2003active,oros2007let}, and enabling personalized guidance and encouragement from more experienced peers~\cite{adaptive-empathy2022,dittus_2017_cscw}. 
Additionally, social interactions cultivate a sense of belonging, encouraging sustained engagement and motivating ongoing participation and contribution~\cite{tausczik2017, 10.1145/3369255.3369258}.

In recent years, \acp{LLM} have emerged as powerful assistive tools in various tasks, including answering programming questions~\cite{10.1145/3613904.3642377}, helping with debugging~\cite{10.1145/3661145}, and often serving as convenient alternatives to traditional search-and-explore methods in online communities~\cite{kabir2024}.
Despite effectiveness of \acp{LLM}, social interactions such as knowledge exchange and engagement with others’ content within coding communities remain highly meaningful\rev{~\cite{cheng2020, Terragni2022}.}
LLMs, while capable of rapidly generating responses, often provide impersonal, isolated, and static answers that lack context and nuanced understanding~\cite{lee-etal-2024-well}. 
In contrast, coding communities allow users to engage with past conversations, clarify their problems through follow-up questions, and receive tailored responses from peers who share similar challenges\rev{~\cite{cheng2020, Terragni2022,Sengupta2020LearningWC}.}
Additionally, coding communities support critical thinking and engagement in ways LLMs currently do not\rev{~\cite{kabir2024}. }
Online coding communities like Stack Overflow presents multiple answers and public commentary for each query, enabling comparison, clarification, and evaluation of competing perspectives\rev{~\cite{cheng2020, Terragni2022,kabir2024,Sengupta2020LearningWC}. }
This format cultivates argumentation literacy by encouraging users to reason through alternatives\rev{~\cite{gudkova2021developing,Sengupta2020LearningWC}.} 
In contrast, LLM-based tools such as ChatGPT typically provide a single confident response, which may reduce opportunities for deliberation and reinforce passive consumption and delegation among learners\rev{~\cite{yang2024can}.}
Moreover, while LLMs are helpful, they often suffer from hallucinations --- generating plausible yet factually incorrect responses without proper citations~\cite{sun2023contrastive,rawte-etal-2023-troubling,kabir2024}. 
These inaccuracies can significantly undermine user trust, especially in programming contexts where precision and reliability are critical~\cite{wang2024investigating}. 

\begin{figure}[t]
    \centering
    \includegraphics[width=0.5\textwidth]{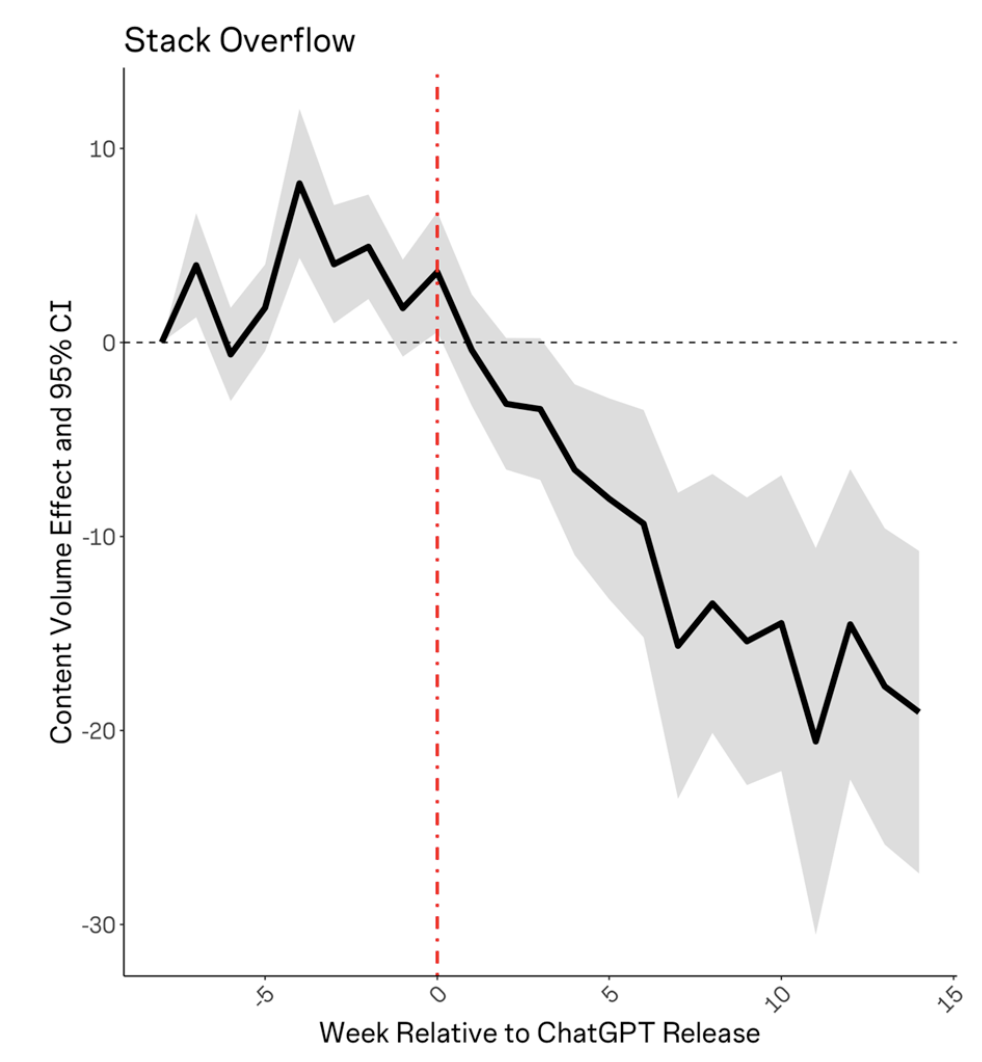}
    \vspace{-10pt}
    \caption{Effects of ChatGPT over time (by week) on Stack Overflow question volumes per topic, with ChatGPT released on November 30th, 2022. Adapted from ~\cite{burtch2024}.}
    \label{fig:fig0}
    \vspace{-10pt}
\end{figure}

On the other hand, the increasing reliance on \ac{LLM}-based tools like ChatGPT endangers active participation to online coding communities~\cite{burtch2024}. 
As illustrated in Figure~\ref{fig:fig0}, the number of questions posted per topic on Stack Overflow has significantly decreased since the release of ChatGPT. 
As more people turn to \ac{LLM}-based tools for quick answers instead of engaging in communities, not only are fewer contributions made, but the incentives to contribute are also diminished~\cite{burtch2024}. 
Recent research within the \ac{CSCW} and \ac{HCI} community further emphasizes this trend, highlighting how dependency on LLMs can diminish essential social interactions among students, negatively impacting peer relationships and collaborative learning opportunities~\cite{park_promise_2024,flathmann2024_cscw}. 

\rev{Given the widespread adoption and practical benefits of LLM through chatbots, it is not feasible to abandon them; instead, there is an opportunity to redesign them in ways that better support community engagement.}
Therefore the question we pose is : \textbf{\textit{How can we design LLM-based chatbots to augment, rather than displace community participation in online coding communities?}}
\rev{In particular, we focus on designing a chatbot that redirects users' attention back toward the community rather than keeping it within the chatbot itself.}
In this paper, we introduce a novel design paradigm called \emph{\concept{}}, which embeds social learning dynamics into AI responses by surfacing user-generated content and \features{} from online coding communities like Kaggle. At the core of this paradigm is a \ac{RAG} mechanism that grounds AI responses in relevant community notebook posts. This grounding not only reduces response hallucinations~\cite{lewis2020,shuster-etal-2021-retrieval-augmentation}, but also ensures that responses reflect diverse, real-world problem-solving approaches from the community. 
Building on top of this, our design displays previews of source posts along with rich social cues such as author profiles, vote counts, view counts, and more. 
These features are deliberately chosen to increase social transparency and social presence in the interaction.
Drawing from social transparency theory~\cite{ehsan2021,stuart2012social}, we aim to make the social context of knowledge creation visible, allowing users to see who contributed the information and how the community engaged with it.
\rev{This visibility supports trust calibration, informed decision making, and social interaction~\cite{ehsan2021,stuart2012social,erickson2000}.}
Similarly, social presence theory~\cite{oh2018systematic,socialpresencecscw2} informs our inclusion of identity cues and interaction signals that help users feel connected to real community members, rather than interacting with an isolated AI.
By embedding these theories into the interface design, we aim to not only encourage users to critically assess AI responses, but also support \rev{engagement} with the online community.
In particular, the design promotes ``lurking'' behavior --- the passive consumption of community content --- which research has shown to be a valuable early step toward active participation in online communities\rev{~\cite{nonnecke2000lurker,preece2004top,lurking2006}}.
At the same time, it reduces the friction traditionally associated with browsing and searching Kaggle threads.

To evaluate our design, we instantiate it in \sys{}, a chatbot that operationalizes the concept of \concept{} for the Kaggle forum. 
We conducted two user studies to evaluate the effectiveness and explore the design space of \concept{}. 
In the first study, we compared \sys{} with three baseline approaches --- \ac{RAG}-baseline, GPT-4o, and traditional browsing --- using a within-subjects lab study with 28 participants. 
The results indicate that learners leveraged the community-enriched previews to navigate toward Kaggle community posts, providing direct evidence of \sys{}'s effectiveness in encouraging engagement and participation in online coding communities.
This design significantly improves the perceived reliability of the content compared to presenting the same generated content without community context.

Building on these findings, we conducted a second design exploration study with 12 participants to examine how varying levels of community integration --- ranging from minimal links to full summaries with \features{} --- influence users’ trust and community engagement. 
Participants responded most positively to designs that paired AI's answers with rich visual previews of relevant posts and visible social signals, such as author profiles and vote counts. Designs that aggregated and summarized the community perspectives were deemed especially effective in enhancing trust and helping users feel more confident in their understanding. By making community knowledge and interactions visible through these features, users were able to \rev{engage} with peer insights and explore content more efficiently --- thereby fostering self-directed learning.
%
Together, the two studies highlight the importance of integrating community knowledge in AI systems to facilitate socially-informed learning.

In summary, our work makes the following contributions:
\begin{itemize}
    \item \emph{Community-Enriched AI}, a design paradigm that enhances AI chatbots with user-contributed content from online coding communities, explicitly displaying previews of sources alongside rich \features{}.
    \item \sys{}, a chatbot integrates Community-Enriched AI design paradigm to assist data science learners on the Kaggle platform.
    \item An empirical evaluation of \sys{} confirms that the Community-Enriched AI design can enable the first step toward improving community participation, enhance users’ trust in model responses, and effectively support learners with data science tasks.
    \item Provides design insights on integrating varying levels of social features into Community-Enriched AI systems to facilitate users' trust in AI and \rev{toward meaningful engagement} with the community. 
\end{itemize}

\section{Related Work}

\subsection{Community Engagement in Online Coding Communities}
Learning programming and software development is often a community-driven endeavor where online coding communities and forums play a central role in facilitating knowledge sharing, learning through collective discussions, deliberation and sharing of feedback~\cite{sengupta_2020, Dondio_Shaheen_2020, cheng2020}. 
Online coding communities, including question-and-answer forums like Stack Overflow and domain-specific platforms with open innovation competitions, like Kaggle, are widely used by both novices and experienced developers for problem-solving, sharing code and solutions, and engaging in informal learning through unstructured activities and social interactions~\cite{cheng2020, Terragni2022,Sengupta2020LearningWC}. 
Learners participate in online coding communities not only to improve their programming and data science skills but also to engage in socially grounded practices like sharing solutions, commenting on alternative approaches, and collaboratively building community knowledge~\cite{cheng2020,tausczik2017,wang2013}. 

Participation in online communities has been explored through different theoretical lenses~\cite{preece2009reader, lave_legitimate_2001}. For example, the reader-to-leader framework describe participants as \textit{readers} who browse and search for content, \textit{contributors} who post and rate content, \textit{collaborators} who work together to improve the community, and \textit{leaders} who promote participation and mentor novices~\cite{preece2009reader}. A common characteristic across these frameworks is that most contributions are made by a minority of participants. Prior work suggests that over 90\% of participants are not active contributors~\cite{carron-arthur_describing_2014}, indicating that the majority of users in online coding communities are readers. Understanding how to design features that enhance readers’ social presence is crucial, as this can foster greater community engagement and potentially encourage readers to transition into active roles such as contributors. To achieve this, it is essential to consider how online communities promote their presence and encourage participation of potential members~\cite{kraut_building_2011}.

While participating in online coding communities are supported in multiple ways, there are still barriers to participation that might limit learning. Specifically, as communities grow, it becomes increasingly difficult for learners to screen relevant content~\cite{stansfield_exploring_2016,vermette_cscw,gilmer_cscw}. Another limitation is that when learners post a question in an online coding community, the response time can vary, and some questions may not receive a response at all~\cite{asaduzzaman2013,fang2023_cscw}. 

To address these barriers, \ac{CSCW} and \ac{HCI} researchers have explored ways to pair novices with experienced community mentors, helping novices adhere to community cultural norms when asking questions, which may lead to quicker responses~\cite{ford2018}. Other researchers have collected and processed Kaggle notebooks as datasets to facilitate research on code metrics in Jupyter notebooks~\cite{mostafavi2024}. However, these methods do not utilize community resources to provide timely and contextually relevant responses to learners’ questions.

\subsection{Automatic Help Seeking in Code Learning}
Before the \acp{LLM} era, search engines and information retrieval tools were crucial for programming tasks, providing access to tutorials, documentation, and community discussions~\cite{brandt2009two, hoffmann2007assieme}. 
In the HCI community, Hoffmann et al.~\cite{hoffmann2007assieme} introduced Assieme, a search interface that consolidated distributed programming resources such as API documentation, sample code, and explanations. 
By resolving implicit references, Assieme enabled programmers to find better solutions with fewer queries compared to general-purpose search engines.
Brandt et al.~\cite{brandt2009two} further examined programmers' use of online resources, identifying opportunistic learning behaviors such as just-in-time learning, clarifying existing knowledge, and memory aids. 
These studies highlighted the limitations of traditional search tools in addressing programming learners' contextual needs --- such as integrating fragmented resources and supporting task-specific information seeking~\cite{brandt2009two, hoffmann2007assieme}.

With the advancement of \ac{LLM}, AI-assisted tools have shown great potential in providing on-demand help to programming learners.
Recent work in \ac{CSCW} and \ac{HCI} has explored the design and deployment of \ac{LLM}-based tools tailored to programming education, which can be broadly categorized into two directions: tools that serve as instructional scaffolds, and those that function as problem-solving aids. Firstly, LLMs have been used as instructional scaffolds to support conceptual understanding and cognitive development. For example, Kazemitabaar et al.\cite{kazemitabaar2024codeaid} introduced a programming assistant that responds to conceptual questions and generates pseudocode with explanations to help learners reason about program structure. Extending this direction, Ma et al.\cite{ma_2025} proposed DBox, an interactive system that scaffolds algorithmic problem-solving through learner-LLM co-decomposition. By guiding students to incrementally construct a step tree and aligning their logic with real-time LLM feedback, DBox fosters critical thinking, cognitive engagement, and independent problem-solving. Expanding this focus to collaborative contexts, Yan et al.~\cite{yan2025llm} examined the role of LLMs in collaborative programming among middle school students. Their findings suggest that LLM-enhanced collaboration significantly improves computational thinking and reduces cognitive load.

LLMs have also been deployed as problem-solving aids to assist learners with concrete coding tasks. Yang et al.\cite{yang2024debugging} examined help-seeking behaviors while learners interacted with a pedagogically designed chatbot to debug code, identifying strategies students use to engage with AI support. Complementing this perspective, Prather et al.\cite{prather2024widening} investigated novices’ experiences with LLMs and found that learners with foundational knowledge benefited from LLM-supported coding. Moving beyond reactive help, Chen et al.~\cite{chen2025} explored proactive LLM-based assistants that offer context-aware programming support without explicit user prompts. Their study showed that proactive suggestions integrated into the coding environment improved user productivity and experience, while also highlighting design trade-offs in mixed-initiative human-AI collaboration. 



\subsection{Challenges LLMs Present to Learners and Existing Solutions}
Despite their benefits, LLM-based tools pose certain challenges for learners.
Firstly, \ac{LLM}s are prone to \say{hallucinations}~\cite{rawte-etal-2023-troubling} which can propagate low quality or incorrect knowledge to the generated content~\cite{kabir2024}.
As identified by Park et al.~\cite{park_promise_2024} \emph{hallucinations}, among other algorithmic problems (e.g., algorithmic bias, lack of transparency and interpretability), are a critical limitation of \ac{LLM}s which impedes productive learning. As a result, learners need a mechanism to validate whether AI-generated responses are trustworthy. While HCI researchers have proposed the design and adoption of specialized \ac{LLM}s for education and the incorporation of fact-checking algorithms~\cite{park_promise_2024}, another approach is to provide source information relevant to the query for the generation system, a concept central to \ac{RAG}~\cite{lewis2020}.
The \ac{RAG} system integrates a retriever and a generator to enhance question-answering. The retriever retrieves relevant information from a data collection, which is then used as context by the generator to produce answers with reduced hallucination~\cite{lewis2020,shuster-etal-2021-retrieval-augmentation}. However, researchers have not yet explored whether or how providing sources for learners is beneficial. A recent HCI study show that incorporating community-curated experiences helps build trust in AI code generation tools~\cite{cheng2024}. Although their tools were only tested on a small, simulated community and focused solely on code generation, their findings inspired us to build \sys{}, which provides users with real community-curated insights for data science tasks.

Secondly, reliance on \ac{LLM}s for accessing information on coding and software development can also reduce user participation in online coding communities, thereby hindering the dynamic exchange of knowledge and limiting opportunities for learning from these communities. Recent research has found a decline in daily web traffic on Stack Overflow by approximately 1 million individuals per day following ChatGPT’s release~\cite{burtch2024}. It has also been observed, question posting volumes per topic on Stack Overflow have markedly declined since ChatGPT’s introduction --- showing the disconnect of learners from online coding communities. Figure~\ref{fig:fig0} from their paper illustrates the changes in question posting volumes per topic on Stack Overflow. 
In educational contexts, studies have shown that the use of LLMs can reduce social interactions among learners creating isolated learning experiences, thus emphasizing the importance of designing systems that encourage peer engagement alongside AI assistance~\cite{park_promise_2024}. Moreover, an over-reliance on LLMs for quick, convenient answers can undermine learners’ opportunities for in-depth learning and the development of critical thinking skills.
In programming, where the fostering of these skills is most acute, the effects of using GenAI have been shown to compound the difficulties in learning coding, especially for struggling learners ~\cite{prather2024widening, park_promise_2024}. Reduced participation in online coding communities further diminishes learners’ opportunities for social learning, narrowing the scope for developing deeper knowledge and problem-solving abilities through engagement with the communities.
\ac{RAG} models have the potential to provide sources to connect the learners back to online community. 
Recent efforts such as Social-RAG~\cite{wang2025socialrag} have begun exploring how retrieval from prior group interactions can socially ground AI generated messages. 
However, HCI researchers have not yet explored the effects of such approach on engaging participation in online coding communities.

This paper bridges online coding communities and LLMs by integrating community-driven features to overcome limitations of current LLM tools.
To do so, we explore designing a Community-Enriched AI system --- \sys{} --- that leverages \ac{RAG} and social design features to reconnect users with peer-contributed content and foster community engagement for learning. 
We contribute to the understanding of socially enriched AI-assisted programming by evaluating the benefits of a Community-Enriched AI design compared to traditional information retrieval methods or general chatbots lacking such integration, providing insights relevant to broader educational ecosystems.

\section{\concept{}: Design Paradigm and Instantiation}

We introduce \concept{}, a design paradigm that integrates the social context of community-generated knowledge into LLM-based chatbots. 
To demonstrate this idea, we instantiate it in \sys{}, a \ac{RAG}-based chatbot that grounds AI-generated responses to the posts of the Kaggle community.
\subsection{Design Motivations for \concept{}}
\subsubsection{Provide Opportunities to Engage with Other Learners}\label{sec:motivation_1}
Data science learners use online spaces to network, learn, and share resources and passions for topics --- interacting continuously and fostering communities of practice~\cite{shrestha2021, wenger_etal_2002}. As an online platform for the data science community, Kaggle provides a space for learners to practice data science skills and engage with others through well-established competitions, community knowledge-building activities such as public code sharing, and social Q\&A-based discussions~\cite{cheng2020}.
The design of our \concept{} aims to create a congruence between the AI user and the social dynamics of the Kaggle community, enabling learners to complement their individual learning through \rev{social engagement}.

Our design is informed by principles from social transparency theory~\cite{stuart2012social} and social presence theory~\cite{socialpresencecscw2}, as applied to AI system design~\cite{ehsan2021}. 
Social transparency highlights the value of making the social context of knowledge creation visible to enhance collaboration and cooperative behaviors~\cite{stuart2012social, ehsan2021}. 
Social presence, on the other hand, highlights the importance of fostering a sense of connection and human interaction within a system, where cues of social presence can increase community participation~\cite{oh2018systematic, socialpresencecscw2}. 
Drawing from these two perspective, our design intends to display peer-generated content with \features{} that can encourage users, like passive readers~\cite{preece2009reader} to engage with the community and also support the content creators by boosting the visibility of their profiles. 



\subsubsection{\rev{Toward Meaningful Engagement} through Community-Guided Information Foraging}\label{sec:motivation_2}
Traditional help-seeking approaches, such as using Google Search, can be time-consuming and overwhelming for learners to find relevant resources that meet their specific needs~\cite{zerhoudi2025,gwizdka2010distribution}, especially when tackling new or domain-specific competitions. 
An effective system should streamline this process by not only helping users find relevant content but also supporting meaningful engagement with that content to promote deeper exploration and critical thinking~\cite{maiti2025}.
By leveraging a retrieval-based approach grounded in community curated knowledge space, our design aims to optimally enable this process for learners by providing them with contextually appropriate resources, and also building a segue for them to engage with community knowledge for thoughtful exploration and learning on the topic.

\subsubsection{\rev{Calibrate} User Trust in LLM-based Assistants}\label{sec:motivation_3}
When learners seek support from LLM-based assistants like ChatGPT, they can encounter incorrect or vague responses that do not fully address their needs~\cite{kabir2024}. 
\rev{Effective LLM-based assistants should aim to reduce hallucinations by grounding responses in community-sourced information~\cite{fan2024survey, ayala-bechard-2024-reducing}, helping learners calibrate an appropriate level of trust in the provided responses~\cite{10.1145/3742413.3789154}.}
Furthermore, learners' interactions with LLMs for programming and data science learning often occur within broader social and community contexts, rather than in isolation.
Research in CSCW and HCI has emphasized the need to contextualize users' trust in AI tools within the broader socio-organizational environment, where relevant collective knowledge and community norms are embedded~\cite{ehsan2021, chen2024}. 
Our system builds on prior research showing that \ac{RAG} can help mitigate hallucinations \cite{lewis2020, shuster-etal-2021-retrieval-augmentation}. In developing \sys{}, we adopt these established mechanisms as a technical foundation, while our contribution lies in extending them with social features that surface community knowledge and cues. 
\rev{Rather than aiming to simply enhance users' confidence in AI-generated answers, \sys{} is designed to support more appropriate trust calibration by enabling learners to cross-check AI-generated content with community-derived examples.} 
Reducing hallucination is not the focus of our study; however, by using \ac{RAG}, our system inherits its benefits and may help lessen hallucination-related issues in practice.

\subsection{The Design of \sys{}} \label{sec:design_chatcommunity}

We developed \sys{}, a \concept{} chatbot that provides answers grounded in community-sourced content \cite{fan2024survey, ayala-bechard-2024-reducing} from the Kaggle platform, previewed with rich community metadata.
As shown in Figure \ref{fig:fig1}, we adopted a UI design that aligns with popular LLM chatbots, where users can submit queries through a text box (Figure \ref{fig:fig1}.A). 
Responses are streamed in real time, include syntax-aware formatting and session memory, with a ``New Chat'' option to reset.

\subsubsection{Iterative Design Process}
The development of \sys{} followed an iterative design process. Initially, the system provided post previews with links to source content. 
A pilot study (Appendix~\ref{sec:pilot}) revealed the need to enhance user engagement, as participants expressed frustration over the lack of community context and recognition for contributors. 
This led us to include rich \features{} --- metadata such as author profiles, author names, publish dates, vote counts, view counts, comment counts, and post titles --- to credit authors and foster interaction.
Based on additional feedback, we added an advanced search panel with ranking options (relevance, vote count, and view count), and allowed users to adjust the number of posts used to generate each response. 

\subsubsection{Source Document Panel}
The source document panel (Figure~\ref{fig:fig1}.C) displays a preview of the source post along with rich \features{} including author profiles, vote counts, view counts, comment counts, publish date and more.
The content preview helps users quickly grasp the relevant information from the source post at a glance, enabling them to decide whether the post is worth interacting with. 
Vote counts, view counts, and comment counts indicate the post’s popularity, while the publish date helps users assess its recency. 
The author’s profile picture fosters a sense of interacting with content created by real people, enhancing engagement with community peers. 
Together, these rich \features{} present what other users select and interact with, fostering a sense of social presence among peers~\cite{oh2018systematic}. 
By making interactions visible, the system engages even passive participants, such as lurkers, by exposing them to valuable content and community activities --- the first step toward deeper engagement~\cite{preece2009reader}. 
This visibility not only facilitates effective communication and collaboration but also enhances the system’s trustworthiness, as highlighted by social transparency theory~\cite{ehsan2021,stuart2012social} and social presence theory~\cite{oh2018systematic,socialpresencecscw2}. 
Overall, the source document panel supports all three design motivations (Sections~\ref{sec:motivation_1}–\ref{sec:motivation_3}).

\subsubsection{Advanced Search Panel}
To balance popularity with relevance, we designed an advanced search panel (Figure~\ref{fig:fig1}.D) with ranking modes -- relevance, votes, and views. 
Relevance uses semantic similarity; votes and views integrate popularity signals.
This avoids over-amplifying only the most popular content and gives users control to surface a broader range of posts. 
To grant users greater control, the advanced search panel allows them to adjust the number of posts each response is based on, ranging from 1 to 10.
Overall, the advanced search panel helps users efficiently locate relevant and reliable posts, supporting the second design motivation (Section~\ref{sec:motivation_2}).

\subsection{Technical Architecture}\label{sec:pipeline}
\begin{figure*}[h]
    \centering
    \includegraphics[width=\textwidth]{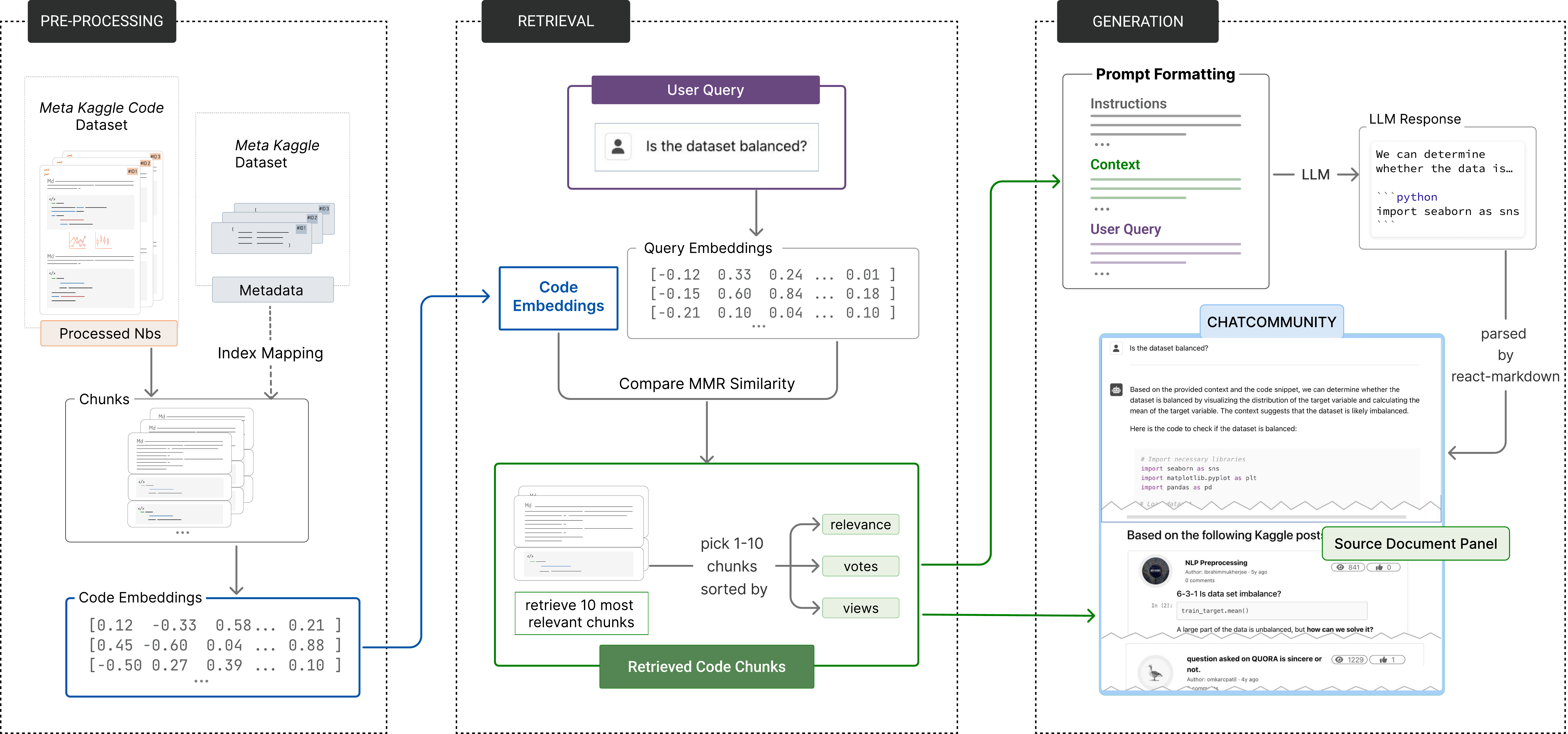}
    \vspace{-10pt}
    \caption{
    RAG Pipeline: The process begins with preprocessing the Kaggle notebooks from the Meta Kaggle Code dataset to extract non-empty, Python-written Jupyter notebooks, which are then organized by individual competitions. Each notebook is further divided into several chunks, and these chunks are linked with the corresponding metadata from the Meta Kaggle dataset through indexing.
Next, the chunks are passed through an encoder to generate embeddings. The embeddings, along with the corresponding chunk content, are stored in ChromaDB. When a user inputs a query, the query is converted into embeddings, and our system searches for and ranks relevant chunks using user-selected sorting methods (relevance, votes, or views).
The retrieved chunks, combined with the user’s query, are then formatted into a complete prompt and passed to LLMs to generate a coherent response. These responses are parsed by our system frontend using the react-markdown package and displayed as system outputs. The retrieved chunks, together with their metadata, are also shown in the source document panel below the system response.
    }
    \label{fig:figure3}
\end{figure*}

Our \ac{RAG} system is specifically designed for online coding communities like Kaggle, leveraging its public notebook posts to generate responses. 
The \ac{RAG} pipeline is illustrated in Figure \ref{fig:figure3}, it comprises two key modules: the retriever and the generator. This section introduces the data source and provides a brief overview of both modules.

\subsubsection{Data Source and Preprocessing}
Our data sources are the \say{Meta Kaggle Code}~\cite{metakagglecode} dataset (2023) and the \say{Meta Kaggle}~\cite{kaggleMetaKaggle} dataset (2016). Both datasets are officially released by Kaggle and are updated weekly until May 7, 2024.
The Meta Kaggle dataset contains Kaggle's public data on competitions, users, submission scores, and kernels, where the Meta Kaggle Code dataset contains the raw source code of the submissions.
These datasets, spanning 2015 to 2024, include 4.83 million code files~\cite{kaggleMetaKaggle,metakagglecode}, with 4.36 million Jupyter notebooks forming the focus of our study.
More details are shown in Appendix~\ref{sec:dataset_detail}.

We first group the code files by competition and filter files from a selection of four competitions (Appendix~\ref{sec:competition}). 
These competitions cover diverse topics, including natural language processing, computer vision, medical data analysis, and business data analysis, ensuring our technical pipeline works effectively with various types of data science competition notebooks. 

Although we do not process all the code files due to resource limitations, our pipeline can be easily applied to different competitions by simply providing the competition identification number. 
We extract non-empty code files in Jupyter Notebook format that were written in Python. 
In summary, we processed 37,895 code files and obtained 36,945 valid Python-written Jupyter Notebook files, containing 849,900 code cells and 269,855 markdown cells. Our data collection adheres strictly to Kaggle's policies. More details are shown in Appendix~\ref{sec:dataset_process}.

\subsubsection{Retriever and Generator Modules}  
We construct the Kaggle post database by dividing each notebook into smaller chunks \(\mathbf{c_i}\).
\rev{Each chunk is formed by grouping consecutive markdown cells followed by the consecutive code cells that immediately follow them.}
This chunking ensures efficient retrieval and balances prompt length. Each chunk is embedded into a vector representation using the \texttt{text-embedding-ada-002} model~\cite{openai_new_2024}, with embeddings stored in a ChromaDB vector database~\cite{chromadb24}.
\rev{As the Meta Kaggle dataset stores social signals (e.g., view counts) for the corresponding notebooks in the Meta Kaggle Code dataset, we use the notebook ID to link each chunk to its respective notebook's metadata, thereby attaching the relevant social signals to the chunks. This ensures that every chunk retains the social context of its original notebook.}

When a user submits a query, we retrieve the top 10\footnote{The choice of the top 10 is common practice in information retrieval~\cite{jones1999phrasier}. } most relevant chunks using the Maximal Marginal Relevance (MMR)~\cite{carbonell1998use} scoring method, which balances relevance and diversity by reducing redundancy. These chunks are ranked based on one of three user-specified criteria: relevance (MMR score), view count, or vote count. The top \(N\) chunks, as specified by the user, are then presented with contextual information about their respective Kaggle posts. The default search is based on relevance, giving newer or low-vote posts a chance to display if relevant to the query.

Finally, the retrieved chunks and the user query are combined into a structured prompt, which is processed by the GPT-4o model~\cite{gpt4o2024}\footnote{
We choose the GPT-4o model as our task focuses on general data science tasks. At the time of conducting the study, GPT-4o was proven to be one of the best-performing models in data science code generation and mathematics tasks~\cite{ouyang2025dscodebench,huang2024olympicarena} and hence deemed as an appropriate choice.}.
The model's responses are rendered using the react-markdown~\cite{npmjsReactmarkdown} package and displayed in a streaming, typewriter manner. For code cells, the raw source is wrapped in fenced Markdown (triple backticks with a language tag such as python) and parsed by react-markdown; these code blocks are then rendered with syntax highlighting using the react-syntax-highlighter~\cite{npmjsReactsyntaxhighlighter} package. Markdown cells are passed directly to react-markdown.
Retrieved chunks are displayed as source previews in the source document panel. For each chunk, we look up its entry in the metadata dataset using its ID and show the associated information, including the post's URL, which is rendered as a hyperlink that directs the user to the source post.
Further details are available in Appendix \ref{sec:RAG-details}.

\subsubsection{Implementation}\label{sec:implementation}
We implemented \sys{} as a web application for easy access.
The backend is built with Flask and integrates our \ac{RAG} pipeline. We used LangChain~\cite{langchain} to handle API requests to GPT-4o model and ChromaDB~\cite{chromadb24} to store embeddings for retrieval within the \ac{RAG} pipeline. The backend receives messages from the frontend, queries the pipeline, and returns the model’s responses together with the retrieved chunks and their corresponding metadata.
The frontend was developed from scratch using React. To render responses, we employed the react-markdown~\cite{npmjsReactmarkdown} package, which directly parses and displays LLM outputs in Markdown format, including both text and code blocks.\footnote{The pipeline is open-sourced at: https://github.com/ETH-PEACH-Lab/ChatCommunity.}

\section{Study 1: User Study of the Community-Enriched AI Design}
To investigate the effectiveness of the Community-Enriched AI design paradigm,  we first conducted a user study to address the following research question:

\begin{itemize}
    \item[] \textbf{RQ1.} \textit{How does the Community-Enriched AI design influence users’ engagement with online coding communities, their perceived reliability, and their performance on data science tasks? }
\end{itemize}
We employed a within-subjects study with four experimental conditions, involving 28 data science learners who had prior experience with Kaggle or similar coding competitions. Participants were instructed to solve four tasks within a single Kaggle competition, using four different assistance methods (as shown in Figure \ref{fig:fig1}), each constituting a distinct experimental condition.


\subsection{Four Assistance Methods}\label{sec:three_design}
To understand how learners perceive the benefits of \concept{} design, we implemented three chatbot variants for comparison, using the same study apparatus (as shown in Figure~\ref{fig:fig1}.).
As a baseline condition, we also included a setting where users directly utilize the Kaggle platform without chatbot assistance.
During the study, we referred to the conditions by their names (Alpha, Beta, Gamma, Delta) when introducing the session to participants to minimize any bias associated with the names. 
\begin{itemize}
    \item \textbf{Community-Enriched AI (Alpha)}: Community-Enriched AI incorporates all the designs and functions mentioned in Section~\ref{sec:design_chatcommunity}. It is equipped with our specially designed \ac{RAG} model and implements the \concept{} design paradigm;
    \item \textbf{\ac{RAG}-baseline (Beta)}: \ac{RAG}-baseline utilizes the same \ac{RAG} model as \concept{}, but without the source document panel nor the advanced search panel. By omitting these, we can evaluate whether the presence of these elements impacts users’ performance on data science tasks and their perceptions of the system;
    \item \textbf{GPT-4o-based mode (Gamma)}: In this setting, we leverage the GPT-4o API to generate responses to user queries. GPT-4o relies on its pre-trained language model to produce answers without directly referencing specific posts from coding communities like Kaggle.
    \item \textbf{Browsing mode (Delta)}: In this setting, we provide participants with a link to the current Kaggle competition page, allowing them to freely search and browse content on the platform.
\end{itemize}

\subsection{Participants and Recruitment}
We conducted a power analysis using G*Power \cite{faul2009statistical} to estimate the required sample size for our within-subjects study design. 
Based on pilot studies (Appendix~\ref{sec:pilot}), we assumed an effect size of $f = 0.7$, with a significance level of $\alpha = 0.05$ and power ($1 - \beta$) of 0.81, resulting in a required sample size of 27 participants.
After securing ethical approval, we recruited 29 graduate and undergraduate STEM students through social media platforms (e.g., LinkedIn). 
Qualified participants are self-identified as experienced data science learners, including 27 graduate students and 2 senior undergraduate, with 20 affiliated with the author’s institution, and 27 residing in the author’s country. 
Ages ranged from 19 to 29, with 18 males, 10 females, and 1 undisclosed. One participant who failed to complete questionnaires and appeared distracted was excluded, leaving 28 valid data points. More details are in Table~\ref{tab:tab7}. 
\subsection{Study Protocol}

All participants signed the consent form before the study. During the study, participants first received a 10-minute introduction on how to set up the study environment and use \sys{}. 
Participants opened a Google Colab notebook with task descriptions alongside the \sys{} system. 
We asked them to place the notebook on the left half of the screen and \sys{} on the right half. 
To ensure participants fully understood how to use \sys{}, we demonstrated each function and then gave them time to explore the system on their own. We began the study only after participants indicated that they were familiar with the system. During this time, we encouraged them to ask any questions they had about using \sys{}.

Participants then completed four data science tasks, each with a time limit of 13 minutes and representing a typical challenge in data science projects.
For each task, participants answered two questions using one of the four assisting methods.  
\rev{The first task required participants to write a few lines of Python code to inspect the task-related dataset and answer questions about its properties, whereas the remaining three tasks asked participants to respond to conceptual questions without writing code. Details about the task questions are listed in Appendix Table~\ref{tab:grading_criteria}.}
We used a Balanced Latin Square design~\cite{bradley1958complete} to arrange the order of assisting methods for each participant, minimizing the order effects of the methods.


After each task, participants completed a Likert-scale post-task questionnaire to evaluate the assisting method. 
After completing all tasks, participants filled out a post-session questionnaire ranking the methods by usefulness and reliability. We also conducted 10-minute semi-structured exit interviews and collected completed notebooks for analysis.

The study lasted 60–90 minutes and was conducted virtually via Zoom. 
Participants used the chatbots in their own browsers, completed tasks in a provided Google Colab notebook. 
Each received 47 USD as compensation.

\subsection{Study Task}
We selected the \say{Quora Insincere Questions Classification} competition~\cite{kaggleQuoraInsincere} for our user study due to its straightforward nature as a binary classification problem.
We also ensured that none of the participants had previously worked on this competition through the screening survey.
Grounded in this competition, we designed a set of questions tailored to key stages in data science workflow: data loading, data preprocessing, model development, and model evaluation. 
We set a time limit of 13 minutes for each task, based on our pilot study, where we found that participants could complete the tasks within this time.
We designed two data-oriented or decision-making questions crucial for solving data science problems in each task, as shown in Appendix~\ref{sec:grade_distribution}.
These questions were reviewed by two data science experts and reflect common challenges encountered in data science projects. 

\subsection{Data Collection and Analysis}
This study collected data from multiple sources, including screening questionnaires, notebook results, post-task and post-session questionnaires, exit interviews, observation notes, and system usage logs. We report the significance levels ($p$) and test statistics ($Z$) for each statistical test in this section and Section~\ref{sec:result}. 

\subsubsection{Task Performance} 
We evaluate the task performance through the notebook grades and task completion time.
Notebook grades were assigned on a scale from 0 to 3, where 3 indicates a fully correct, complete, and relevant answer.
Two researchers independently graded each notebook's results using predefined evaluation criteria (Appendix~\ref{sec:grading_criteria}). 
Out of 224 grades, there were five discrepancies between the researchers.
The two researchers discussed these differences and reached a consensus on the final grades. 

Regarding notebook grades, the linear regression analysis showed no learning effect between tasks ($p$ = 0.53)~\cite{lazar2017research}. The Shapiro–Wilk test indicated grades were not normally distributed. We then used Friedman tests to analyze the effects of task type and assisting method.
The analysis revealed significant effects of both task type ($p$<0.01, $Z$=13.34), and assisting method ($p$<0.01, $Z$=33.87), on notebook grades. We normalized grades within each task using Z-scores~\cite{abdi2010normalizing} to minimize the effects of task type. As participants in the Alpha condition had higher average grades, we performed a post-hoc analysis using a one-sided Wilcoxon signed-rank test with Bonferroni correction, comparing Alpha to other conditions.

Regarding task completion time, the Shapiro–Wilk test showed they were not normally distributed.
Friedman tests were conducted and revealed significant effects of assisting method ($p$ < 0.01, $Z$ = 24.66), but no significant effect of task type ($p$ = 0.49, $Z$ = 2.40),  on task completion time. Therefore, we conducted a post-hoc analysis using a two-sided Wilcoxon signed-rank test with Bonferroni correction on task completion time, comparing Alpha to other conditions.
The analysis results of both notebook grades and task completion time are displayed in Table~\ref{tab:anova_table2}. Detailed grading results for each question are provided in Appendix~\ref{sec:grade_distribution}.

\subsubsection{Post-Task Questionnaires and Post-Session Questionnaires}
The post-task questionnaires used 7-point Likert scales to evaluate assisting methods on various attributes.
The post-session questionnaire contains two ranking questions, with results shown in Figure~\ref{fig:figure5}.
%
The Shapiro–Wilk test showed none of the post-task questionnaire scores were normally distributed. Friedman tests revealed no significant effect of task type on any score, while the assisting method had a significant effect across all scores. As Alpha consistently had a higher mean score than the other conditions, we performed one-sided post-hoc Wilcoxon signed-rank tests with Bonferroni correction to compare Alpha with other conditions. Results are in Table~\ref{tab:anova_table1}.


\subsubsection{Exit Interview}
We conducted exit interviews after the session, recording 316 minutes of video to gather additional feedback on the assisting methods.
We transcribed and proofread participants’ answers and the recordings to produce the interview transcripts. We use the interview transcripts as anecdotal evidence to support our quantitative findings. Therefore, we used \textit{in vivo}~\cite{saldana_coding_2013} coding to analyze the interviews and attune ourselves to the users' perspectives on the different designs.

\subsubsection{Observation Notes and Usage Logs}
During each session, a researcher took observational notes while the system recorded participants’ queries, responses, and function usage. 
These data were used to capture user behavior and provide insights into how participants interacted with the different methods. The screen recordings were also transcribed and used to review participant behaviors. 

\section{Study 1: Findings}\label{sec:result}

\subsection{How Does Community-Enriched AI Encourage Engagement in Online Coding Communities?}
We define participation in an online coding community as behaviors such as viewing, voting, commenting, contacting peers, or contributing posts to the community~\cite{wang2012understanding,hellman2022}.
Beyond increased viewing through chatbot previews, most participants (22) clicked on the previews to read the original Kaggle posts in Alpha, leading to a total of 77 posts read, with each participant clicking on more than two posts on average. 
We also observed that two participants gave votes to three posts that they found useful.
We asked participants for their reasons for clicking on post previews, which included verifying the system response (10), viewing more details of the post (6), and interacting with the post author (4). This suggests that the source document panel design drives specific engagement behaviors by encouraging participants to verify AI outputs for trust, pursue detailed insights, and initiate direct interactions with post authors, highlighting its role in \rev{supporting meaningful community engagement}.
In terms of prompt iteration, participants in the Alpha condition submitted the highest number of input prompts (M = 5.25, SD = 2.84), followed by those in the Gamma (M = 4.04, SD = 2.22) and Beta (M = 3.96, SD = 1.75) conditions. This pattern is consistent with our observational notes: participants in the Alpha condition frequently engaged with post previews and original posts to gather additional information and used the advanced search panel to refine their retrieval strategies. On average, each participant in the Alpha condition used the advanced search panel more than once.
Additional behavioral statistics are available in Appendix~\ref{sec:user_behavior}.

We further probed into how this design affects participants as potential contributors to the posts.
Nearly all participants (27) preferred having their past and future posts linked by the chatbot system, with most (22) already having experience contributing posts to online communities.
Participants recognized that having their posts linked by the chatbot would lead to more people reading their posts (12) and provide help to more people (9), potentially increasing engagement in the community.
Participants also highlighted conditions for linking their posts, including only linking their \inlinequote{open-sourced posts} (P1, P26) and ensuring that the \inlinequote{author’s name is displayed} (P8, P15).
Most participants (22) affirmed that having their public posts explicitly referenced by the chatbot, potentially increasing views, likes, or comments from peers, would motivate them to contribute more to the community.
Participants felt that this design would allow more people to use and interact with their posts, enabling real communication, encouraging them to contribute more.
As P4 mentioned: \inlinequote{thanks to the chatbot, more people would know [about my post]; so yeah, it would definitely encourage me}, and as P7 said: \inlinequote{Its kind of like you not talking to the wall, you are talking to people, so its better}.

\subsection{How Do Participants Perceive the Reliability of Community-Enriched AI?}
We probe into perceived reliability by analyzing the results of post-session questionnaires, post-task questionnaires, and interviews. 
As shown in Figure~\ref{fig:figure5}, all participants ranked Alpha as the most (18) or second most (10) reliable method.

When comparing Alpha with Beta, Alpha has significantly higher scores in \say{information provided is reliable} ($p$ < 0.01, $Z$ = 1.32) and \say{felt confident using the method} ($p$ < 0.01, $Z$ = 0.03) according to post-hoc analysis. 
There are 21 participants mentioned that displaying the source post previews along with the social features made the system's responses more reliable, as it \inlinequote{helps to check whether it gives the correct answer} (P1) and \inlinequote{could see what Kaggle posts were used by the chatbot} (P16). 
\rev{Our observation notes further support this. For example, during the task with Alpha, P1 first read through the model’s response, then examined each post preview, repeatedly switching back and forth between the response and the previews to cross-check the information. Similarly, P16 read the model's response, clicked on a post preview with a high view count (798) to open the original Kaggle post, compared its content with the chatbot's answer, and subsequently enriched their own task response.}
This result demonstrates that, even with identical generated responses in the two conditions, Community-Enriched AI fosters perceived reliability by improving social transparency \cite{ehsan2021,stuart2012social}.

This aligns with participants’ high reliability ratings for Delta, where 14 participants ranked it as the first (8) or second (6) most reliable method. 
Participants appreciated reading content and comments written by ``real people", as mentioned by P11: \inlinequote{Best thing was that there were so many different posts and discussions. Was nice to see real people thinking about the problem}. In the post-hoc analysis, on comparing Alpha with Gamma, we found Alpha has significantly higher score in \say{information provided is reliable}($p$ < 0.01, $Z$ = 2.78) and \say{felt confident using the method}($p$ < 0.01, $Z$ = 0.96). During interviews, 15 participants explicitly mentioned that Gamma was not reliable. Participants described they found Gamma less reliable than Alpha because Gamma did not provide source posts as references, and its responses were sometimes not relevant and off-topic.
As P2 mentioned: \inlinequote{no reference make it less reliable}, and as P15 said: \inlinequote{gives weird answers. It is a little bit off-topic.}

\begin{table*}[h]
\centering
\caption{Notebook grades and task completion time analysis. We used Friedman tests to assess the effects of task type and condition on notebook grades and task completion time. Significant effects were found for task type (p < 0.01) and condition (p < 0.01) on grades, and method on task completion time (p < 0.01), but no effect of task on task completion time (p = 0.49). Post-hoc one-sided Wilcoxon-Signed-Rank test with Bonferroni correction on Z-scores, which were normalized to account for task-related effects, revealed that participants using Alpha condition had significantly higher grades than Gamma and Delta. While the post-hoc two-sided Wilcoxon Signed-Rank test with Bonferroni correction on task completion time found that participants in the Delta condition had significantly higher task completion times than those in the Alpha condition.}
\resizebox{\textwidth}{!}
{%
\begin{tabular}{p{2cm} p{1.2cm} p{0.2cm} p{0.3cm} p{0.3cm} p{0.7cm} p{0.5cm} c}
\toprule
\textbf{Measurement} & \textbf{Cond.} & \textbf{N} & \textbf{M} & \textbf{SD} & \textbf{p1} & \textbf{p2} & \textbf{Value} \\
\midrule
\multirow{4}{3cm}{Notebook \\grades} 
& Alpha & 28 & 5.61 & 0.63 &\multirow{4}{*}{\textbf{0.00**}} & - & \multirow{4}{*}{\includegraphics[width=6.6cm,height=1.5cm]{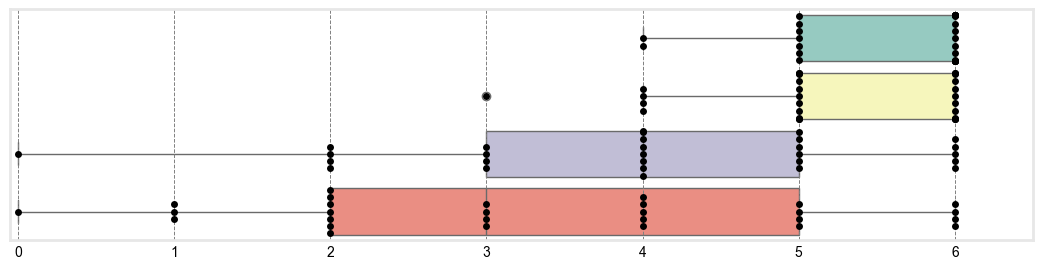}} \\
& Beta & 28 & 5.21 & 0.83 &  & 0.08 & \\
& Gamma & 28 & 4.00 & 1.52 &  & \textbf{0.00**} & \\
& Delta & 28 & 3.32 & 1.74 &  & \textbf{0.00**} & \\
\midrule
\multirow{4}{3cm}{Completion \\time (mins)} 
& Alpha & 28 & 9.52 & 2.50 & \multirow{4}{*}{\textbf{0.00*}} &  - & \multirow{4}{*}{\includegraphics[width=6.6cm,height=1.5cm]{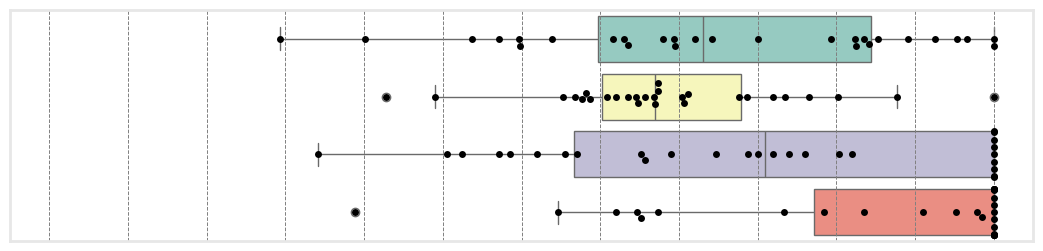}} \\
& Beta & 28 & 8.88 & 1.60 &  & 0.45  & \\
& Gamma & 28 & 9.95 & 2.66 &  & 1.00  & \\
& Delta & 28 & 11.57 & 2.25 & & \textbf{0.01**}  & \\

\bottomrule
\end{tabular}%
}
\label{tab:anova_table2}
\end{table*}

\subsection{How Does Community-Enriched AI Improve Participants’ Performance in Data Science Tasks?}

We probe into user performance by analyzing notebook grades and task completion time as presented in Table~\ref{tab:anova_table2}.
For task completion time, the post-hoc analysis revealed that students using Alpha took significantly less time than Delta ($p$ = 0.01, $Z$ = -3.36). Most participants (26) mentioned that Delta is ineffective and time-consuming, as it is hard to find posts relevant to their task.
For notebook grades, participants in the Alpha condition had the highest average grade across the four tasks.
The results of post-hoc analysis revealed that students using Alpha had significantly higher grades than Gamma ($p$ < 0.01, $Z$ = 4.01) and Delta ($p$ < 0.01, $Z$ = 4.42). 
When analyzing task types (detailed in Appendix~\ref{sec:grade_distribution}), we found that participants performed similarly on data-oriented questions (e.g., Is the training dataset balanced?), but their performance varied significantly on decision-making questions (e.g., Which model should we use and why?). 
This indicates that Community-Enriched AI might be particularly effective in supporting higher-order cognitive tasks, which require synthesizing information, evaluating options, and justifying choices, compared to data-oriented questions that involve basic comprehension and retrieval of factual information.

In addition, we probe into participants' perceived learning. As shown in Table~\ref{tab:anova_table1}.
Alpha has significantly higher score in \say{helpful in learning coding} comparing to other conditions. 
While our evidence does not directly prove its effectiveness for learning, the system’s ability to encourage reading more posts may lead to potential learning gains~\cite{manarin2019read}.
\newcommand{\figurewidth}{5cm}
\begin{table*}[h]
\centering
\caption{Analysis of the post-task questionnaire results: We conducted two Friedman tests to examine the effects of task type and assisting method on the post-task questionnaire responses. The analysis revealed no significant effect of task type on any question, while assisting method had a significant effect across all questions (p-values are reported in column p1). Given that the Alpha condition consistently had a higher mean score than the other conditions, we performed one-sided post-hoc Wilcoxon signed-rank tests with Bonferroni correction for each question, comparing Alpha with the other conditions (p-values are reported in column p2).}
\resizebox{\textwidth}{!}
{%
\begin{tabular}{p{3.2cm} p{1.2cm} p{0.3cm} p{0.3cm} p{0.3cm} p{0.8cm} p{0.8cm} c}
\toprule
\textbf{Statement} & \textbf{Cond.} & \textbf{N} & \textbf{M} & \textbf{SD} & \textbf{p1} & \textbf{p2} & \textbf{Agreement: 1 to 7} \\
\midrule
\multirow{4}{3cm}{The current assisting method is easy to use.} 
& Alpha & 28 & 6.43 & 0.63 & \multirow{4}{*}{\textbf{0.00**}} & - & \multirow{4}{*}{\includegraphics[width=\figurewidth]{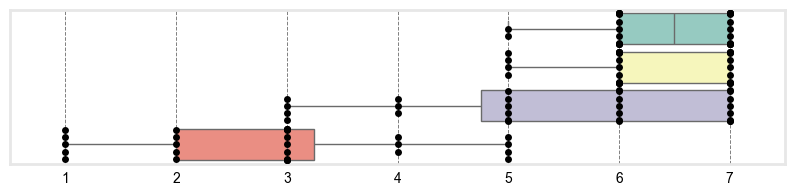}} \\
& Beta & 28 & 6.25 & 0.70 & & 0.40 & \\
& Gamma & 28 & 5.43 & 1.40 & & \textbf{0.00**} & \\
& Delta & 28 & 2.86 & 1.27 & & \textbf{0.00**} & \\
\midrule
\multirow{4}{3cm}{The current assisting method is easy to learn.} 
& Alpha & 28 & 6.61 & 0.50 & \multirow{4}{*}{\textbf{0.00**}} & - & \multirow{4}{*}{\includegraphics[width=\figurewidth]{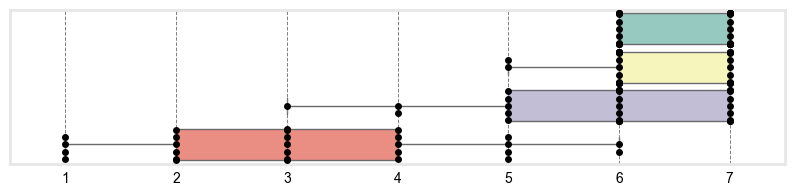}} \\
& Beta & 28 & 6.32 & 0.61 & & 0.14 & \\
& Gamma & 28 & 5.89 & 1.07 & & \textbf{0.01**} & \\
& Delta & 28 & 3.18 & 1.49 & & \textbf{0.00**} & \\
\midrule
\multirow{4}{3cm}{The current assisting method is enjoyable to use.} 
& Alpha & 28 & 6.32 & 0.77 & \multirow{4}{*}{\textbf{0.00**}} & - & \multirow{4}{*}{\includegraphics[width=\figurewidth]{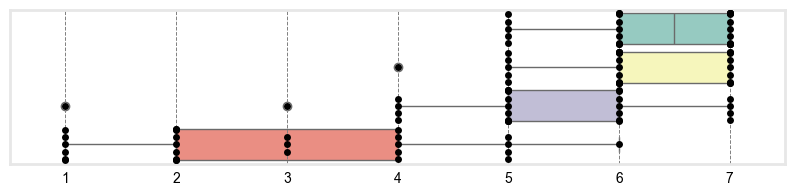}} \\
& Beta & 28 & 6.07 & 0.81 & & 0.31 & \\
& Gamma & 28 & 5.18 & 1.31 & & \textbf{0.00**} & \\
& Delta & 28 & 2.82 & 1.52 & & \textbf{0.00**} & \\
\midrule
\multirow{4}{3cm}{The current assisting method is helpful for me in learning coding.} 
& Alpha & 28 & 6.29 & 1.05 & \multirow{4}{*}{\textbf{0.00**}} & - &  \multirow{4}{*}{\includegraphics[width=\figurewidth]{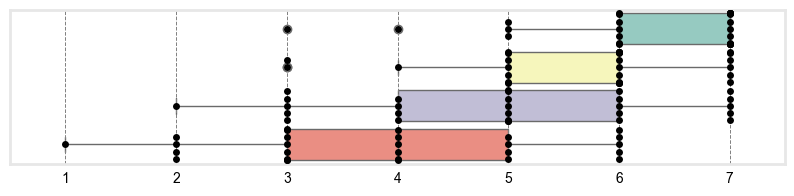}} \\
& Beta & 28 & 5.68 & 1.09 & & \textbf{0.01**} & \\
& Gamma & 28 & 4.93 & 1.46 & & \textbf{0.00**} & \\
& Delta & 28 & 3.82 & 1.44 & & \textbf{0.00**} & \\
\midrule
\multirow{4}{3cm}{The current assisting method is helpful for me in solving coding problems.} 
& Alpha & 28 & 6.43 & 0.84 & \multirow{4}{*}{\textbf{0.00**}} & - & \multirow{4}{*}{\includegraphics[width=\figurewidth]{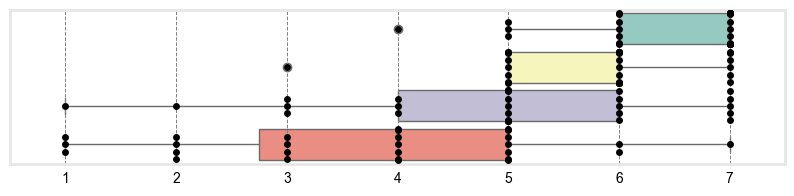}} \\
& Beta & 28 & 5.82 & 0.90 & & \textbf{0.00**} & \\
& Gamma & 28 & 4.96 & 1.55 & & \textbf{0.00**} & \\
& Delta & 28 & 3.75 & 1.60 & & \textbf{0.00**} & \\
\midrule
\multirow{4}{3cm}{The information provided by the current assisting method is reliable.} 
& Alpha & 28 & 6.50 & 0.69 & \multirow{4}{*}{\textbf{0.00**}} & - & \multirow{4}{*}{\includegraphics[width=\figurewidth]{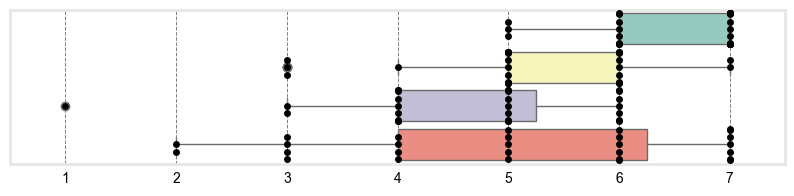}} \\
& Beta & 28 & 5.21 & 1.03 &  & \textbf{0.00**} & \\
& Gamma & 28 & 4.64 & 1.16 &  & \textbf{0.00**} & \\
& Delta & 28 & 5.07 & 1.63 &  & \textbf{0.00**} & \\
\midrule
\multirow{4}{3cm}{I felt very confident using the current assisting method.} 
& Alpha & 28 & 6.43 & 0.74 & \multirow{4}{*}{\textbf{0.00**}} & - & \multirow{4}{*}{\includegraphics[width=\figurewidth]{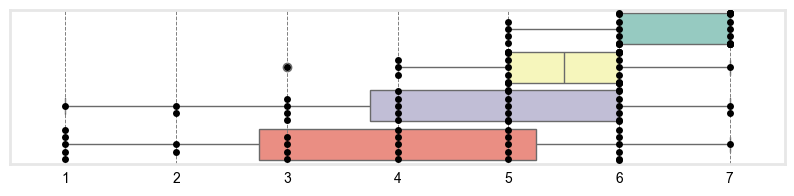}} \\
& Beta & 28 & 5.36 & 0.87 & & \textbf{0.00**} & \\
& Gamma & 28 & 4.57 & 1.55 & & \textbf{0.00**} & \\
& Delta & 28 & 3.89 & 1.87 &  & \textbf{0.00**} & \\
\midrule
\multirow{4}{3cm}{I would like to use this assisting method frequently.} 
& Alpha & 28 & 6.57 & 0.69 & \multirow{4}{*}{\textbf{0.00**}} & - & \multirow{4}{*}{\includegraphics[width=\figurewidth]{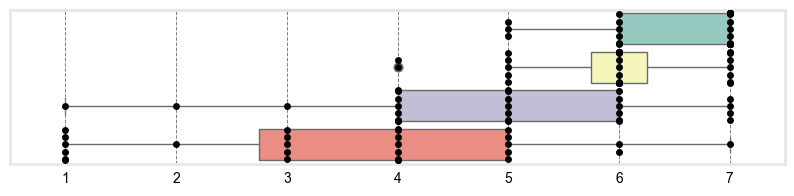}} \\
& Beta & 28 & 5.93 & 0.86 & & \textbf{0.01**} & \\
& Gamma & 28 & 4.89 & 1.47 & & \textbf{0.00**} & \\
& Delta & 28 & 3.54 & 1.71 & & \textbf{0.00**} & \\
\bottomrule
\end{tabular}%
}
\label{tab:anova_table1}
\end{table*}
\subsection{Perceived Usefulness and Usability with Community-Enriched AI}
Lastly, we reported perceived usefulness~\cite{davis1989perceived} and perceived usability~\cite{lewis2018system} by analyzing the results of post-session questionnaires, post-task questionnaires, and exit interviews. 
Most participants (26) ranked Alpha as the most useful method compared to others, as shown in Figure~\ref{fig:figure5}. Table~\ref{tab:anova_table1} presents the statistical results of the post-task questionnaires. In summary, participants find Community-Enriched AI more useful and have higher usability than others.

When comparing Alpha with all other conditions, it scored significantly higher in “helpful for solving coding problems” and “liking to use frequently.” 
Compared to Gamma and Delta, participants rated Alpha significantly better in “easy to use,” “easy to learn,” and “enjoyable to use” according to post-hoc analysis.
Most participants (24) mentioned that Alpha, by displaying the source preview, made the system more useful compared to Beta. Reasons mentioned by participants include that Alpha \inlinequote{provides more evidence... which are good for solving tasks} (P7) and allows users to \inlinequote{get useful solutions from others} (P2). 
That is, participants preferred receiving the original sources from Alpha because the posts provided evidence to support the response and helped them find useful solutions from others. 

Additionally, 20 participants emphasized that providing advanced search panel enhances the system’s usefulness.
P8 said this panel offers more options and makes the information more relevant, while P5 said it increased their trust by including ranking options based on vote and view counts.
During the study, 21 participants used the advanced search panel to switch posts rankings from relevance to votes (16) or views (4) and increase the number of retrieved posts (16).

There are 23 participants mentioned that Gamma is not useful for completing data science tasks comparing to Alpha, given their general (10) and lengthy (7) responses. 
The median word count for Alpha’s responses is 330, compared to 362 for Gamma. While the difference is not significant, Gamma's general responses may increase cognitive load~\cite{jose2011,de2010cognitive}, making the responses feel overly lengthy to users.

Similarly, the Delta condition may also impose a high cognitive load by requiring users to browse and interpret large amounts of community posts manually~\cite{jose2011,de2010cognitive}.
Most participants (26) mentioned Delta is ineffective and time-consuming. P24 said it’s hard to find relevant information with Delta. We observed that four participants gave up Delta midway through the task and attempted to complete it on their own. While participants are allowed to use any function including the search engine in Kaggle, and most (24) participants have used the search engine, they did not find it useful, as P1 mentioned \inlinequote{It is the least efficient method I can find. Because you need to go through multiple layers of searching...You need to look into it... it's quite tiring.} 

\section{Study 2: Design Exploration Study of Levels of Community Features Integration in AI Chatbots}\label{sec:explore_result}
\rev{
Through our Study 1, we demonstrate the significance of \concept{} as a design paradigm and its measurable impact on users' decisions. We show that incorporating community contributions to AI chatbots' responses encourages users to engage with the community and also enhances their trust and perceived reliability on the responses. However, the specific design elements of these community-enriched features and the extent to which they should be integrated into the design of socially-enhanced AI systems remain underexplored.
To better understand the design space of pertinent social features for Community-Enriched AI systems, we conduct Study 2 and pose the following research question:}

\begin{itemize}
    \item[] \textbf{RQ2.} \textit{Which community-enriched design attributes do users value most for fostering trust in AI chatbot responses and encouraging engagement with the community? }
\end{itemize}
  
\subsection{Study Protocol}
To address RQ2 we conducted a qualitative design exploration study examining four levels of community integration in AI chatbots --- ranging from minimal cues such as titles and links, to previews and social features, and finally to rich summaries reflecting community consensus and disagreement. This study investigates how the varying degrees of community features shape user preferences, perceptions and their behavior, particularly in fostering trust in AI chatbot responses and promoting \rev{community engagement}.

\subsubsection{Participants}
The last twelve participants from the previous study (P17–P28; see Table~\ref{tab:tab7}) took part in this study as an extended session following a short break (5–10 minutes) after Study 1. 
The Study 2 session lasted 30–40 minutes.
As a token of appreciation for completing both sessions (total duration: 100–120 minutes), participants received a total of \$53 USD, which exceeds the local minimum wage.

\subsubsection{Procedure}
In this study, each session began with an introduction to four design variations, all of which were implemented as high-fidelity prototypes in Figma using a Wizard-of-Oz approach~\cite{dahlback1993wizard,porcheron2021pulling}. 
This was followed by a task in which participants used each variation to answer a data science question about the same Kaggle competition used in the previous user study: ``Which model should we use for this competition?''. Following the task, we conducted interviews to understand users' perceptions of the various attributes of the different designs, their perceived benefits and drawbacks for each design variants. 

\begin{figure*}[h]
    \centering
    \includegraphics[width=\textwidth]{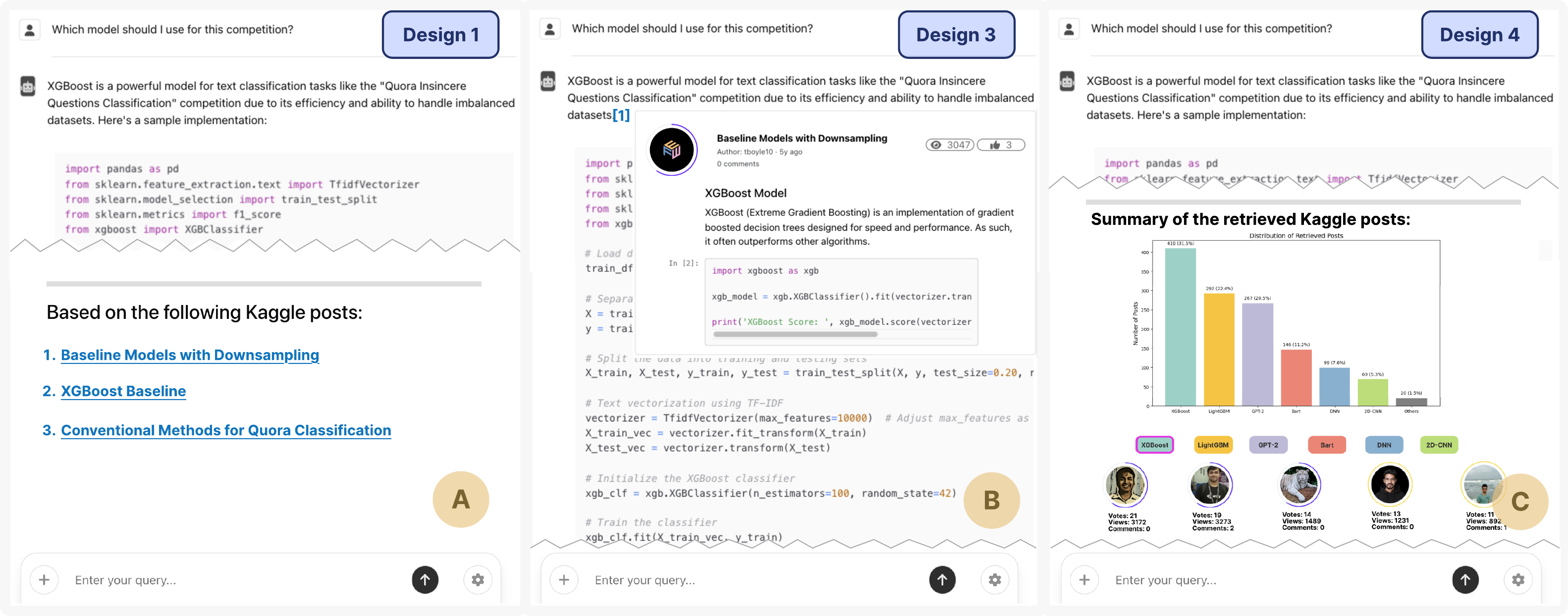}
    \vspace{-10pt}
    \caption{An overview of design variations: Design 1: Vanilla Link, Design 3: Community-Enriched Inline and Design 4: Community-Enriched Summary. Design 2: Community-Enriched Preview is the same as Figure~\ref{fig:fig1}.Alpha.
    }
    \Description{}
    \label{fig:figure6}
\end{figure*}
\subsubsection{Design Variations\label{sec:design_variations}}
All four design variations presented the same textual chatbot response and shared the same ranking of relevant posts. Designs 1–3 used the same set of retrieved posts, while Design 4 included additional posts to support each solution. The main differences across the variations lie in how the retrieved posts and associated \features{} are presented to the user.

\begin{itemize}
    \item  \textbf{Design 1: Vanilla Link.} The Vanilla Link design (Figure \ref{fig:figure6}A) displays the source post as a simple hyperlink to the post title, providing a straightforward and minimal design.
    \item   \textbf{Design 2: Community-Enriched Preview.} The Community-Enriched Preview is the same design as Figure \ref{fig:fig1} Alpha from the previous study, providing post previews along with social features like author profiles, view and vote counts.
    \item  \textbf{Design 3: Community-Enriched Inline.} The Community-Enriched Inline (Figure \ref{fig:figure6}B) provides sentence-level references with clickable links to community-enriched previews, offering in-context support. This design was inspired by participant feedback in Study 1, where P6, P9, and P14 emphasized a need for fine-grained inline references to better pinpoint details in the original post.
    \item  \textbf{Design 4: Community-Enriched Summary.} The Community-Enriched Summary (Figure \ref{fig:figure6}C) summarizes retrieved posts, allowing users to view the distribution of solutions across all relevant posts. This design is inspired by prior research~\cite{zhang2019empirical,zhu2022empirical} showing that summarizing online coding community discussions can help users compare alternative solutions more effectively. The top 10 posts, ranked by relevance, vote count, or view count as chosen by the user, are displayed alongside the \features{} such as user profiles, vote counts, view counts, and comment counts.
\end{itemize}

\subsubsection{Data Analysis}
A total of $306$ minutes of interview was recorded. The interviews were transcribed and two authors of the paper analyzed the transcripts using Reflexive Thematic Analysis~\cite{braun_one_2020}. The goal of the analysis was to understand which community features, and to what degrees of their integration, shaped users' trust and confidence in the AI chatbots' responses, influenced their community engagement, and provided them with opportunities to engage with the content.
The first two authors independently coded five randomly sampled transcripts. 
They then discussed the codes and reached an agreement level of 97\%~\cite{miles2020qualitative}. 
The remaining transcripts were divided equally at random and coded independently by the two authors. 
The resulting codes were jointly discussed and organized into relevant themes.

\section{Study 2: Findings}
The coding process from our thematic analysis yielded 167 unique codes. These were iteratively interpreted and arranged into four themes. In the following sections, we present each theme with illustrative quotes from participants.

\subsection{Fostering trust and reliability through encapsulating community-enriched content}
One of the key design attributes emerging from our design variations, and preferred by all users, is the integration of community knowledge with AI-generated assistance --- presenting both sources and responses together. Participants reflected that \textbf{viewing a comprehensive community enriched context as an overview of posts alongside the chatbot’s responses} helped them gain reliability on its response. As P20 describes, \inlinequote{I think [in Design 2] since knowing who answered this question and when they answer and how many people view it, this information makes me trust the answer more. This is because I can directly see the content of the answer, and also since the source and answer appear together.} In addition to fostering trust, having access to this contextual information also helped users assess the relevance of the answer to their needs. As P25 noted: \inlinequote{From the preview I could know whether it's related to what I want to find. If it's interesting I would click on that and go to the original post.} 

Our findings also reveal variation in users’ preferences regarding how this encapsulated information should be displayed. All users (12) preferred minimizing the effort required to access community-enriched information by \textbf{visualizing it as a preview of relevant posts}, rather than embedding it as inline references within the chatbot’s response. This preview-based approach was perceived as more usable for navigating information related to the chatbot’s answers. As P18 noted: \inlinequote{Inline reference discourages users from interacting with posts, users need to do extra interaction to see the post…sometimes you just want all the posts below [chatbot’s answers], you can just compare the number of every post, it's quite intuitive and useful for users to check details.} Participants also suggested that these previews could be further enriched by seeing keywords in the preview relevant to their topic. However, some users also perceived value in inline references as depicted in Design 3, in fostering trust in the AI’s responses. This format was seen as similar to citing scientific articles, allowing users to \textbf{inspect sources directly at specific points} in the response. Nonetheless, participants felt that while inline references supported transparency, they did not fully convey source credibility or support reliability as effectively as the preview overview.


\subsection{Fostering trust and reliability through social attributes of community-enriched AI chatbot responses}
As reflected in the previous theme, users expressed that integrating community-enriched information alongside chatbot responses enhances both the perceived reliability and relevance of the answers. This integration allows them to make social inferences by drawing on the diverse attributes present in community knowledge which also contributes to calibrating users’ trust in the answers of the AI chatbot. Specifically, social design features from our design variations: the Community-Enriched Posts Preview feature from Design 2 and the Community-Enriched Summary feature from Design 4 contributed to users' reliability and trust in AI's responses. Below we present the social design attributes from these two design variations that users deemed useful and necessary for fostering trust and reliability in AI responses.

First, the users emphasized the importance of \textbf{author identity transparency}, such as displaying the author’s name, profile information, and image on the posts along with the response of the AI chatbot. These elements acted as cues, helping users trust and  feel confident in the AI response. From the context of social transparency in AI systems~\cite{ehsan2021}, this social design attribute introduces human elements of decision making in the system that helps shape people’s perception of and trust in the AI’s response. As P18 reflected, \inlinequote{[In Design 2]...the post is created by real people. GPT only provides you with articles, you don't have access to the author name, profile picture, you don't know who created the content, making you not confident enough about these contents.}  

Second, the other social design attribute that contributes to building reliability on the AI’s response is the visibility of community engagement with the posts, achieved through \textbf{interaction transparency}~\cite{stuart2012social} --- specifically by displaying interaction metrics such as votes, views and clicks on the post previews as implemented in Design 2. Of these interaction attributes, in particular most users emphasized the importance of votes, followed by metrics on views on the posts as a cues to build a socially-situated significant perception of the AI’s answer. P19 explained, \inlinequote{The like and view feature definitely encourages reliability [on the chatbot’s response], it is saying many other people are doing it, other people agree with it.} Providing information about community engagement allows users to infer the significance of the answer, fostering transitive trust in the AI system by trusting the community’s endorsement~\cite{ehsan2021}. Beyond ratings, metrics such as the number of comments on a post preview also helped build transitive trust. As P28 noted, \inlinequote{Comment is important. More comments means more people review this article.}

Third, extending the design attributes that foster transitive trust in the AI’s responses, our feature of summarizing retrieved posts and presenting the distribution of solutions across all relevant posts was perceived as further reinforcing users’ trust and confidence in the chatbot’s answers. This design attribute not only makes the technological grounding of the AI’s answers visible by anchoring the summary in a ``scientific way'' --- but also allows users to see an \textbf{aggregated view of community engagement with the collective information}, infusing the human element to complement AI’s reasoning~\cite{ehsan2021}. As P27 shared: \inlinequote{It makes me feel the decision of clicking links is statistics-based, science-based. I like that.} By providing an overview of how solutions are being used by others in the community, the design helps users visualize the human element embedded in decision-making, reinforcing trust through social heuristics like social endorsement,  As P21 explained: \inlinequote{[Design 4] shows the comparison from the whole picture --- which one is the most used one. We have a comparison, and it provides more confidence regarding reliability.}, or by inferring relevance from others' action, as mentioned by P28, \inlinequote{More users use this method, I will try to think about why more users use it. Because more users use this, this may be more suitable for this problem.}


\subsection{Fostering social engagement with the community}
Findings from our interviews further show that while providing community-enriched information about source posts fosters users’ trust in AI responses, it also serves as a bridge between the AI and online coding communities by offering affordances for users to engage with the original community posts.

Firstly, participants expressed that providing the community enriched information about the source posts as \textbf{direct visual previews} are more enticing for users to engage with the posts. P17 expressed \inlinequote{making preview more directly embedded is better for community discussion.}
Secondly, the \textbf{social design attributes} of these post previews --- highlighted in the previous theme, including author profiles, post titles, metrics capturing others’ interactions with the posts, and temporal context --- enable users to more readily engage with the posts. 
As P18 noted, \inlinequote{The features [in Design 2] definitely encourage you to interact with the community. You naturally like to click on things here...click the post here, comment on something, vote for him if the post is useful.} 
These social attributes help users find relevance in AI’s answers, which essentially encourages them to read and engage with the posts. And through the transparency of the community attributes as users build transitive trust on AI responses, they are also encouraged to further contribute to the community by expressing their own views and interacting with other community members~\cite{preece2009reader}. 
This progression reflects the stages described in the \textit{Reader-to-Leader Framework}~\cite{preece2009reader}, where users gradually evolve from passive readers to active contributors as they develop trust in the system’s responses.
As P17 said \inlinequote{If I find the answer is good, I can click; If I have more time, I would connect with the author, if I find this post is exactly what I need, I would comment on it.}

However, most of the users also expressed that direct interactions with the community from our chatbot, based on the post previews, are best limited to \textbf{small non-conversational contributions} through likes and votes. This preference stemmed from the recognition that the preview alone does not convey the full context necessary for deeper engagement. As P19 explained, \inlinequote{The chatbot provides only a short preview most of the time, I need to go to the Kaggle page to get the whole context.} To contribute more substantively or to collaborate meaningfully, participants preferred accessing the complete context of the post with its social attributes \textbf{directly within the native online coding community}, where they felt better \textbf{equipped to engage responsibly}. As P27 noted, \inlinequote{I don’t think interaction should happen in the chatbot; the comment should come from someone who has really read the post…it feels unjustified or unfair for me to comment on or vote for a post after only reading a preview.}


\subsection{Fostering social knowledge from community enriched AI responses}

Finally, our findings show that community-enriched features integrated with AI responses not only enhance the interaction but also enabled users for deeper learning through social knowledge-building. 
Through the social feature of summarized information on posts, some users experienced a sense of \textbf{agency in productive exploration and learning}, enabling them to identify preferred models and methods from the rich insights of community consensus.
P24 mentioned \inlinequote{[Design 4] increases my willingness to explore different posts and methods. It shows me there are also other possibilities and makes my curious about how other people using different method. I would like to use \texttt{lightgbm} and see how different methods perform.}
Furthermore, having access to collective knowledge from the community, along with AI’s response, allows users to \textbf{broaden their perspectives and deepen their social learning}. P19 noted \inlinequote{I think the process of looking for a solution is good for you to learn. Here you search for a solution in the community, you are in the product, you also know what other people are doing...broaden my views and get more information.} Participants also emphasized that the tool’s \textbf{community-specific focus} helped them find more ``fine-tuned'' information and fostered a sense of social connectedness with others.

\section{Discussion}
\subsection{Toward Designing Community-Enriched AI Systems}
\subsubsection{Significance of Community-Enriched Design in Perceived Trust}
Our work demonstrates that bridging online coding communities with AI through a Community-Enriched design can significantly enhance users’ trust in AI responses. This trust emerges through two mechanisms: (1) encapsulating community knowledge in AI-generated answers, and (2) displaying social attributes alongside those answers.

Firstly, our Community-Enriched design uses a community-tailored \ac{RAG} model to retrieve relevant posts to meet diverse user needs. 
The advanced search panel further leverages \features{} such as vote count and view count, helping users identify popular and trustworthy content. 
This design improves perceived reliability and usability by building on prior work on \ac{RAG} to reduce hallucinations~\cite{lewis2020,shuster-etal-2021-retrieval-augmentation} and lowering extraneous cognitive load~\cite{de2010cognitive}, compared to GPT-4o and traditional search/browsing methods.

Secondly, by making community knowledge visible through \features{}, our design allows users to anchor their trust in the AI’s responses based on the collective insights of the community. 
When users see that the AI’s answer aligns with content they recognize as trustworthy, they develop transitive trust~\cite{ehsan2021} in the AI — trusting it because it reflects the community’s shared understanding.
By embedding visible cues of peer interactions --- what users select, vote for, or like, as well as how they engage with the community’s shared knowledge --- the interface enables users to draw social inferences from community interactions, leading to more informed trust in the AI~\cite{ehsan2021,stuart2012social}. 
Additionally, incorporating identity transparency in design~\cite{stuart2012social}, such as showing post authors' profile information, provides additional social signals that help users assess about sources and relevance of the AI's answers.
Insights from Study 2 highlight that the level of social attribute visibility also shapes users' trust perceptions.
While simple visual previews of source posts help users ground their judgment, enriching these previews with summaries that capture diverse community perspectives provides greater clarity and reinforces a sense of reliability.

\subsubsection{Supporting Community Engagement and Learning Agency}

In alignment with Cai et al.\cite{cai2024}'s suggestion to raise students' awareness and autonomy in conversational AI, our findings show that Community-Enriched AI not only enhances trust but also supports user engagement and learning agency by surfacing community sourced content that encourages self-directed learning and social participation. 
By presenting previews of related posts along with \features{}, our design encourages users to critically examine AI responses, compare alternative perspectives, and discover more tailored solutions from the community. 
This process fosters learner agency by enabling users to make informed decisions rather than passively accepting AI outputs.

In our second study, participants reported that these community-enriched previews lowered the effort to access additional knowledge and made them more inclined to explore and interact with the original posts. 
Participants shared that they felt comfortable engaging through lightweight actions such as viewing, liking, or voting from within the AI interface but preferred to transition to the full community platform for more substantive contributions. 
This behavior reflects early stages of engagement described in the Reader-to-Leader framework~\cite{preece2009reader}, where users evolve from passive readers to active community participants. 
Our design thus not only supports surface-level engagement but also provides potentials for deeper involvement and responsible knowledge sharing.

Finally, access to community knowledge was also seen as fostering a sense of social connectedness~\cite{park2021_cscw,wang2022}. 
Participants emphasized that seeing how others solved similar problems, and understanding which solutions were most used, helped them develop a broader perspective on problem-solving. 
This collective insight strengthened their learning experience and increased their sense of connectedness within the community. 

\rev{Our findings also highlight a broader question about where engagement should occur, whether it should remain within the chatbot or return to the online community. Participants appreciated the convenience of lightweight interactions such as viewing or voting from within the chatbot, yet most emphasized that conversations with post authors or other users should take place on the original community platform. They noted that the chatbot previews often lacked the full context needed for responsible commenting or collaboration, and that deeper engagement felt more appropriate within the native environment where posts, authors, and discussions were fully visible. This distinction raises important ecological and ethical considerations about the ownership and sustainability of community knowledge in the age of generative AI: when community data are used to train or augment LLMs, who owns the user and who should own the engagement? Design paradigms like \concept{} can serve as mediators in this ecosystem: using community data to enhance AI assistance while still directing recognition, credit, and interaction back to the online communities.}

\subsubsection{Design Takeaways} \label{sec:take-away}
Drawing on insights from our studies, we offer the following design takeaways for future designers and researchers in creating community-supported AI systems.
\begin{itemize}
    \item \textbf{Community-generated content as source previews}: Comprehensive information of sources grounded in community-generated content should be provided as a deictic context to the AI's response. Specifically, these source information should be organized in visual formats like post-previews to allow users easily gain access to relevant community insights and make informed inferences about the AI's answers. 
    
    \item \textbf{Social attributes presented through transparent design}: Alongside source information, AI interface designs should explicitly present the social attributes of referenced content by incorporating socially transparent features --- such as author identity and community interaction transparency~\cite{stuart2012social} to help users build trust in the AI's response. 
    
    \item \textbf{Aggregate community perspectives}: The aggregated engagement data (e.g., votes, views) and community perspectives from source posts should be presented as well-structured summaries to help users make socially informed inferences, identify relevant content, and develop transitive trust in the AI's responses.

    \item \textbf{Embed community information directly within AI responses}: Designs should directly embed visual previews of community posts and their socially transparent design features within AI responses to encourage \rev{user engagement}. These embedded previews enable lightweight, non-conversational interactions with the community, such as likes or votes, allowing users to readily engage with the community content. 
    
    \item \textbf{Enable full-context engagement within the community}: While embedded previews focus on non-conversational community interactions, designs should also support seamless transitions to the original platform for deeper, more responsible contributions --- recognizing users' need for full context when \rev{engaging}.

\end{itemize}

\subsection{Ethical Considerations} \label{ethics}
\subsubsection{Data policy}
When integrating \concept{}, it is essential to ensure that the data used for generating responses adheres to ethical and legal standards.
Thus, the open sourced Meta-data dataset provides us the opportunity to build and evaluate our ideas.
A key feature of \sys{} is displaying sources for each response. 
However, this necessitates careful consideration of the data policies associated with the sources. 
One participant mentioned, \inlinequote{as long as these data [posts] are open-sourced} (P1). 
This indicates that the participant believes the prerequisite for linking posts with the chatbot is that the posts must already be open-sourced.
Future systems should only access publicly available data or content explicitly permitted for public use. 
Using private or restricted data without proper authorization can lead to ethical and legal issues, especially when the generated responses will link to the source content. 
Developers must carefully review data sources to ensure compliance with relevant policies and regulations, such as GDPR~\cite{gdprinfoGeneralData}, ensuring that personal or sensitive data is appropriately anonymized or excluded.

\subsubsection{Control by users}
Empowering users with control over retrieval ranking is a key aspect of maintaining fairness and transparency in \sys{}. 
We provide users with an advanced search panel that offers three objective post retrieval ranking methods: relevance, votes, and views. There are 20 out of 28 participants found that the advanced search panel increased the system’s usefulness.
Designing retrieval methods requires careful consideration to avoid unintended biases~\cite{diaz2008through}. 
For example, if not implemented thoughtfully, certain posts could consistently rank at the top, while others are overlooked, leading to skewed visibility and bias in the information presented. 
To prevent this, future systems building on our work should ensure that the retrieval methods are balanced and do not favor certain types of content disproportionately (e.g., based on business profit).

\subsection{Limitations and Future Work}
Our studies have a number of limitations. For both of our user studies, we designed tasks based on one Kaggle competition~\cite{kaggleQuoraInsincere}, covering stages like data loading, pre-processing, model development, and evaluation. However, it would be valuable to assess \sys{} on other tasks and other platforms. Besides, we focused on data-oriented and decision-making questions, which do not represent all possible data science challenges. However, evidence suggests \ac{RAG} systems improve LLM performance in general question-answering tasks, supporting \sys{}’s generalizability~\cite{lewis2020,siriwardhana-etal-2023-improving}.

In addition, while measuring the perceived reliability, we did not assess the accuracy of the individual AI response. 
And the chunking method may miss contextual information. However, our contribution focuses on the design paradigm of Community-Enriched AI, which remains applicable even to an ideal \ac{RAG} model without the need for chunking.

Although \sys{} improved perceived learning gains, our studies did not assess actual learning outcomes. Therefore, we cannot make definitive claims about its impact on learning effectiveness. However, \sys{} offers better access to Kaggle posts than typical chatbots, 
potentially supporting user learning through active reading of community-generated content~\cite{manarin2019read}.

Finally, although \sys{} is designed to support socially-informed learning, the customization options in the advanced search panel are currently limited to fixed attributes. 
Future work could explore dynamically adjusting results based on user input. 
For example, users might describe their needs in natural language --- \say{I only want posts written in PyTorch, ranked by the number of views}, and the system could retrieve and rank posts accordingly.

\section{Conclusion}

This paper introduces \concept{}, a design paradigm that grounds AI responses in user-contributed content from online coding communities and displays source previews alongside \features{}.
Through two user studies with 28 and 12 data science learners respectively, we demonstrate that a chatbot implementing this design not only enhances perceived system reliability but also encourages \rev{user engagement} with the community and effectively supports learners in solving data science tasks.
Our findings highlight the potential of \concept{} to bridge, rather than replace online coding communities --- offering design implications for building socially grounded AI assistance systems that facilitate productive social learning.
\begin{acks}
This project was made possible by ETH AI Center Doctoral Fellowships to Junling Wang, with partial support from the ETH Zurich Foundation. Additionally, the authors wish to thank the reviewers, members of the PEACH Lab at ETH Zurich, and the participants in the user study.
\end{acks}
\bibliographystyle{ACM-Reference-Format}
\bibliography{ref}
\appendix
\section{Participants’ Demographics}
We present the participants' demographics in Table~\ref{tab:tab7}.
\begin{table*}[h]
    \centering
    \scriptsize
    \caption{Participants’ Demographics: We recruited 29 participants with STEM backgrounds and experience in Kaggle or similar data science competitions. One participant was excluded due to incomplete questionnaires and noticeable distraction during the study, resulting in 28 valid participants.}
    \resizebox{\textwidth}{!}
    {%
    \begin{tabular}{cp{1.5cm}p{3cm}p{0.5cm}p{1cm}p{1.8cm}p{1.6cm}p{1.6cm}}
     \toprule
      \textbf{PID}   & \textbf{Educational Level} & \textbf{Majors} & \textbf{Age} & \textbf{Gender} & \textbf{Experience with Online Coding Communities} & \textbf{Knowledge of Data Science} & \textbf{Knowledge of Python}\\
      \midrule
       1 & Master & Computer Science & 25 & Male & yes & yes & yes\\
       2 & Master & Data Science & 22 & Male & yes & yes & yes\\
       3 & Ph.D & Computer Science & 27 & Male & yes & yes & yes\\
       4 & Master & Data Science & 25 & Female & yes & yes & yes\\
       5 & Master & Mechanical Engineering & 25 & Male & yes & yes & yes\\
       6 & Master & Data Science & 24 & Female & yes & yes & yes\\
       7 & Master & Quantitative Finance & 23 & Male & yes & yes & yes\\
       8 & Master & Mathematical Engineering & 23 & Female & yes & yes & yes\\
       9 & Master & Artificial Intelligence & 28 & Male & yes & yes & yes\\
       10 & Master & Physics & 23 & Male & yes & yes & yes\\
       11 & Master & Data Science & 24 & Female & yes & yes & yes\\
       12 & Master & Computer Science & 23 & Male & yes & yes & yes\\
       13 & Master & Computer Science & 22 & Female & yes & yes & yes\\
       14 & Ph.D & Machine Learning & 25 & Male & yes & yes & yes\\
       15 & Master & Electrical Engineering & 26 & Male & yes & yes & yes\\
       16 & Bachelor & Computer Science & 19 & Prefer not to say & yes & yes & yes\\
       
       17 & Master & Robotics, Systems and Control & 24 & Male & yes & yes & yes\\
       18 & Master & Computational Linguistics & 25 & Male & yes & yes & yes\\
       19 & Master & Computational Linguistics & 23 & Female & yes & yes & yes\\
       20 & Master & Computer Science & 24 & Male & yes & yes & yes\\
       21 & Master & Computational Linguistics & 23 & Male & yes & yes & yes\\
       22 & Master & Data Science & 25 & Female & yes & yes & yes\\
       23 & Ph.D & Computational Social Science & 29 & Female & yes & yes & yes\\
       24 & Bachelor & Mathematics & 27 & Female & yes & yes & yes\\
       25 & Ph.D & Computational Geography & 25 & Male & yes & yes & yes\\
       26 & Master & Computer Science & 25 & Female & yes & yes & yes\\
       27 & Ph.D & Robotics & 26 & Male & yes & yes & yes\\
       28 & Master & Data Science & 25 & Male & yes & yes & yes\\
      \bottomrule       
    \end{tabular}%
    }
    \label{tab:tab7}
\end{table*}
\section{Pilot Study}\label{sec:pilot}
We conducted two pilot studies with two participants: a master's student in data science and a bachelor's student in computer science. We gathered feedback from them on both the system and model design. 

Feedback from the first pilot study highlighted a need to enhance user engagement, as participants expressed frustration over the lack of community context and recognition for contributors. To address this, we integrated rich \features{}, such as author profiles, author names, publish dates, vote counts, view counts, comment counts, and post titles, to credit authors and foster interaction.

In the second pilot study, user feedback emphasized the importance of balancing relevance and popularity in post rankings. Participants highlighted the value of \features{}, such as vote and view counts. In response, we introduced an advanced search panel with three ranking options—relevance, vote count, and view count—and allowed users to adjust the number of posts used to generate responses.

Additionally, participants showed a preference for asking follow-up questions, which informed our decision to design \sys{}’s backend to retain memory of previous conversations. We also observed participants switching between different chatbot systems during testing (Alpha, Beta, and Gamma). To ensure a controlled experiment, we hid the system-switching button and instructed participants to switch systems only when prompted. Further adjustments included repositioning the \say{New Chat} button and refining the backend logic for fetching Kaggle community meta-information.


\begin{table*}[h]
\centering
\caption{Notebook grades distribution}
\scriptsize 
\renewcommand{\arraystretch}{1.3} 
\begin{tabular}{>{\centering\arraybackslash}m{5cm} p{1cm} c c c p{4.5cm}}
\toprule
\textbf{Question} & \textbf{Cond.} & \textbf{N} & \textbf{M} & \textbf{SD} & \textbf{Grade: 0 to 3} \\
\midrule
\multirow{4}{5cm}{1.1 What is the meaning of each attribute in the training and the testing dataset? Are there any missing values in the training dataset?} 
& Alpha & 7 & 2.86 & 0.38 & \multirow{4}{*}{\includegraphics[width=4.5cm]{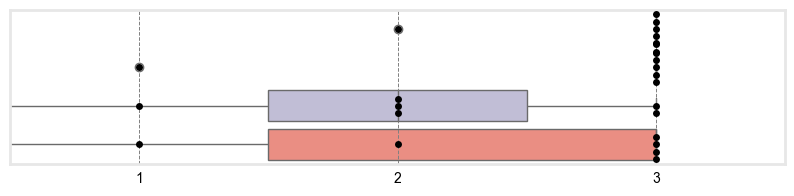}} \\
& Beta & 7 & 2.71 & 0.76 &  \\
& Gamma & 7 & 1.86 & 1.07 &  \\
& Delta & 7 & 2.14 & 1.21 &  \\
\midrule
\multirow{4}{5cm}{1.2 Is the training dataset balanced in terms of class distribution?} 
& Alpha & 7 & 3.00 & 0.00 & \multirow{4}{*}{\includegraphics[width=4.5cm]{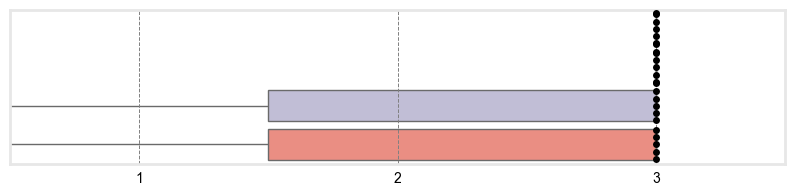}} \\
& Beta & 7 & 3.00 & 0.00 &  \\
& Gamma & 7 & 2.14 & 1.46 &  \\
& Delta & 7 & 2.14 & 1.46 &  \\
\midrule
\multirow{4}{5cm}{2.1 If the training dataset is imbalanced, what methods can we use to handle the imbalanced classes effectively? How might these methods impact the performance of our model?} 
& Alpha & 7 & 3.00 & 0.00 & \multirow{4}{*}{\includegraphics[width=4.5cm]{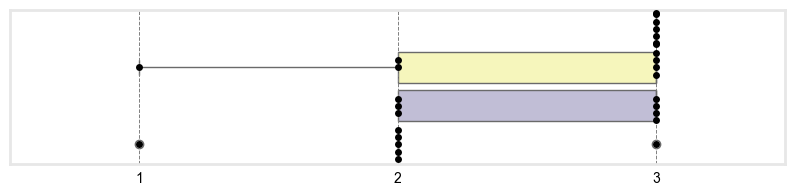}} \\
& Beta & 7 & 2.43 & 0.79 &  \\
& Gamma & 7 & 2.57 & 0.53 &  \\
& Delta & 7 & 2.00 & 0.58 &  \\
\midrule
\multirow{4}{5cm}{2.2 Can we feed the training dataset directly into the model? Explain why?} 
& Alpha & 7 & 2.14 & 0.90 & \multirow{4}{*}{\includegraphics[width=4.5cm]{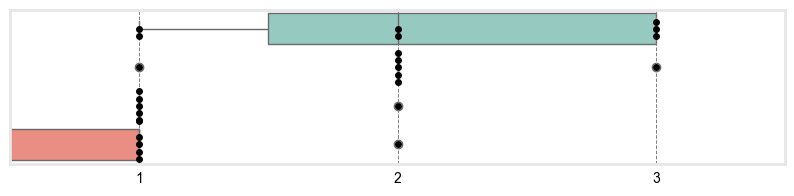}} \\
& Beta & 7 & 2.00 & 0.58 &  \\
& Gamma & 7 & 1.14 & 0.38 &  \\
& Delta & 7 & 0.86 & 0.69 &  \\
\midrule
\multirow{4}{5cm}{3.1 Which embedding method should we use for this competition and why is this method suitable for this competition?} 
& Alpha & 7 & 3.00 & 0.00 & \multirow{4}{*}{\includegraphics[width=4.5cm]{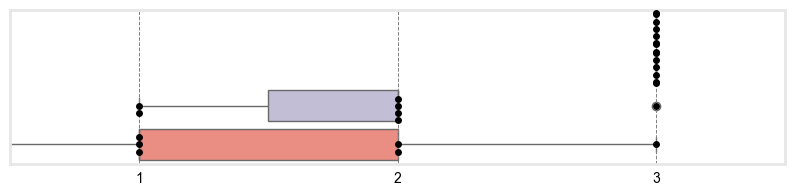}} \\
& Beta & 7 & 3.00 & 0.00 &  \\
& Gamma & 7 & 1.86 & 0.69 &  \\
& Delta & 7 & 1.43 & 0.98 &  \\
\midrule
\multirow{4}{5cm}{3.2 Which model should we use for this competition and why is this model suitable for this competition?} 
& Alpha & 7 & 2.71 & 0.49 & \multirow{4}{*}{\includegraphics[width=4.5cm]{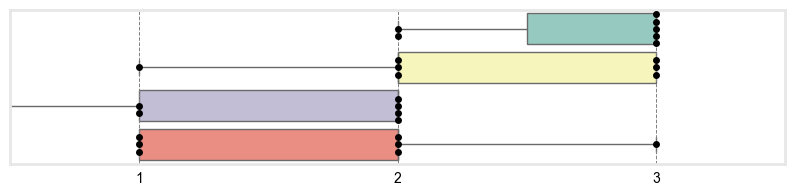}} \\
& Beta & 7 & 2.29 & 0.76 &  \\
& Gamma & 7 & 1.43 & 0.79 &  \\
& Delta & 7 & 1.71 & 0.76 &  \\
\midrule
\multirow{4}{5cm}{4.1 What evaluation metrics should be used to assess the performance of the models in this competition and why?} 
& Alpha & 7 & 2.86 & 0.38 & \multirow{4}{*}{\includegraphics[width=4.5cm]{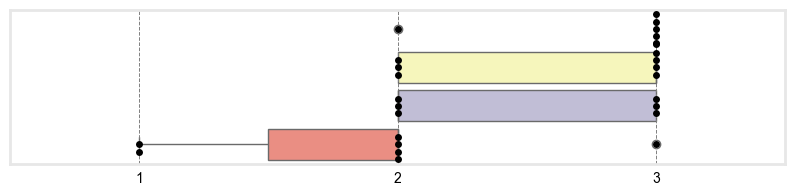}} \\
& Beta & 7 & 2.57 & 0.53 &  \\
& Gamma & 7 & 2.14 & 1.07 &  \\
& Delta & 7 & 1.86 & 0.69 &  \\
\midrule
\multirow{4}{5cm}{4.2 Suppose you are using the Random Forest model as classifier. After initial model training and evaluation, what strategies can further improve the model’s performance?} 
& Alpha & 7 & 2.86 & 0.38 & \multirow{4}{*}{\includegraphics[width=4.5cm]{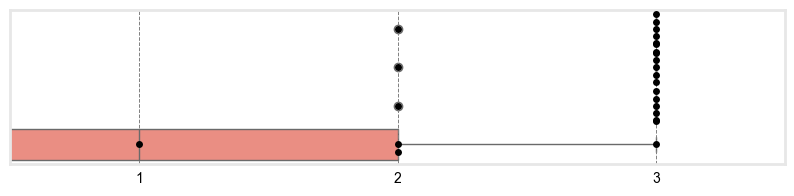}} \\
& Beta & 7 & 2.86 & 0.38 &  \\
& Gamma & 7 & 2.86 & 0.38 &  \\
& Delta & 7 & 1.14 & 1.21 &  \\
\bottomrule
\end{tabular}
\label{tab:anova_table3}
\end{table*}

\section{Grade Distribution for Each Question}\label{sec:grade_distribution}
To better understand the differences in grades, we present the detailed grade distributions for the various questions in Table~\ref{tab:anova_table3}. 
Participants using Alpha achieves the best average grades in seven out of eight questions.  
While all participants performed similarly on data-oriented questions (e.g., Is the training dataset balanced in terms of class distribution?), their performance varied significantly on decision-making questions (e.g., Which model should we use for this competition and why is this model suitable for this competition?).
Specifically, when comparing the answers from Alpha and Gamma on decision-making questions, we found that Gamma tends to provide general suggestions that may not be suitable for the current task. For example, in response to the question: \say{Which model should we use for this competition, and why is it suitable?}, GPT-4o is likely to recommend the BERT model, a transformer-based language model, as a general solution, as reflected in one participant's answer: \inlinequote{I will use BERT model to handle the problem in this competition. It's popular model, easy to finetune and could result a good perfomance in classification task} (P4). The participant chose the BERT model recommended by GPT-4o, despite it being computationally expensive and having a large number of parameters, which makes it less ideal for this binary classification task~\cite{lai2020simple}. In contrast, Community-Enriched AI recommends a suitable model based on posts that have already demonstrated the model’s effectiveness for this task, as reflected in one participant's answer: 

\begin{quote}
    \textit{We can use the simple DNN model. According to the high rated posts, DNN model is able to solve the task with less computational resources than models like transformer. If we want the higher f1-score, we can try the models like transformer or BiLSTM with attention} (P2)
\end{quote} 

This participant chose the simple DNN model which has good flexibility and is able to solve the task with less computational resources. DNN models were used in highly voted posts~\cite{kaggleHandleOverfitting,kaggleAdvancedDeep} and performed well in this competition.
In summary, the Community-Enriched AI design is effective in helping students achieve better performance in specific data science tasks without significantly increasing completion time.

\begin{table}[h]
\centering
\caption{Grading criteria for each question}
\scriptsize
\renewcommand{\arraystretch}{1.3}

\begin{tabular}{>{\raggedright\arraybackslash}m{4cm} >{\raggedright\arraybackslash}m{9cm}}
\toprule
\textbf{Question} & \textbf{Grading Criteria} \\
\midrule
\textbf{1.1 What is the meaning of each attribute in the training and testing dataset? Are there any missing values in the training dataset?} &
\textbf{3 points:} Answer correctly defines \texttt{qid} as "unique identifier" and explains the target as sincere or insincere. And answer no missing values.\newline
\textbf{2 points:} Misses either \texttt{qid} or target definition. Others the same as above.\newline
\textbf{1 point:} Misses both \texttt{qid} and target definition but correctly identifies no missing values.\newline
\textbf{0 points:} Incorrect information on missing values or other factual errors. \\
\midrule
\textbf{1.2 Is the training dataset balanced in terms of class distribution?} &
\textbf{3 points:} Correctly identifies the dataset as imbalanced.\newline
\textbf{0 points:} Incorrectly identifies the dataset as balanced. \\
\midrule
\textbf{2.1 If the training dataset is imbalanced, what methods can we use to handle the imbalance? How might these methods impact model performance?} &
\textbf{3 points:} Identifies 2 or more methods with proper explanations.\newline
\textbf{2 points:} Identifies 1 method with explanation, or 2 methods without proper explanation.\newline
\textbf{1 point:} Identifies 1 method without explanation.\newline
\textbf{0 points:} Incomplete or incorrect answer.\newline
\textbf{Methods to consider:} Resampling (Oversampling, Undersampling), Ensemble Methods, Data Collection, Threshold Adjustment, Class-weight Adjustment, Algorithm Selection, Cost-sensitive Learning. \\
\midrule
\textbf{2.2 Can we feed the training dataset directly into the model? Explain why.} &
\textbf{3 points:} Provides 4 or more preprocessing steps with proper explanations.\newline
\textbf{2 points:} Provides 3 preprocessing steps with explanation.\newline
\textbf{1 point:} Provides only 1 preprocessing step or says "No" without proper explanation.\newline
\textbf{0 points:} Incomplete or incorrect answer.\newline
\textbf{Steps to consider:} Text Cleaning, Tokenization, Text Normalization, Embedding, Padding, Handling Imbalanced Classes, Feature Engineering, Train/Test Split. \\
\midrule
\textbf{3.1 Which embedding method should we use for this competition, and why is it suitable?} &
\textbf{3 points:} Mentions any of the following with proper explanation: Google News, FastText, GloVe, Paragram\_300, Wiki-News, Word2Vec.\newline
\textbf{2 points:} Mentions BERT with a proper explanation, or mentions any of the above methods without proper explanation.\newline
\textbf{1 point:} Mentions BERT without proper explanation or other irrelevant methods.\newline
\textbf{0 points:} Incomplete or incorrect answer. \\
\midrule
\textbf{3.2 Which model should we use for this competition, and why is it suitable?} &
\textbf{3 points:} Identifies models like 2D-CNN, DNN, GPT-2, BART, XGBoost, or LightGBM with proper explanation.\newline
\textbf{2 points:} Mentions BERT, LSTM, or any traditional machine learning models with proper explanation.\newline
\textbf{1 point:} Mentions BERT without explanation, or other models with or without explanation.\newline
\textbf{0 points:} Incomplete or incorrect answer. \\
\midrule
\textbf{4.1 What evaluation metrics should be used to assess the performance of the model?} &
\textbf{3 points:} Identifies 2 or more metrics (e.g., F1 score, AUC, Confusion Matrix) with proper explanation.\newline
\textbf{2 points:} Mentions F1 score, AUC, or Confusion Matrix with proper explanation.\newline
\textbf{1 point:} Mentions 1 metric without proper explanation.\newline
\textbf{0 points:} Incomplete or incorrect answer.\newline
\textbf{Metrics to consider:} F1 Score, AUC, Precision, Recall, Confusion Matrix, Balanced Accuracy. \\
\midrule
\textbf{4.2 What strategies can further improve the performance of the Random Forest model after training and evaluation?} &
\textbf{3 points:} Identifies 3 or more methods with proper explanations.\newline
\textbf{2 points:} Identifies 2 methods with proper explanations.\newline
\textbf{1 point:} Identifies 1 method with or without explanation.\newline
\textbf{0 points:} Incomplete or incorrect answer.\newline
\textbf{Methods to consider:} Hyperparameter Tuning, Feature Engineering, Ensemble Methods, Regularization, Feature Importance Analysis, Data Augmentation, Threshold Tuning, Calibration. \\
\bottomrule
\end{tabular}

\label{tab:grading_criteria}
\end{table}
\section{Grading Criteria}\label{sec:grading_criteria}
The grading criteria were defined by two researchers and reviewed by two data science experts. The grading criteria grade each result based on correctness, relevance, and completeness. The detailed grading criteria are shown in Table~\ref{tab:grading_criteria}.

\section{Results of Post-Session Questionnaire}
\begin{figure*}[h]
    \centering
    \includegraphics[width=\textwidth]{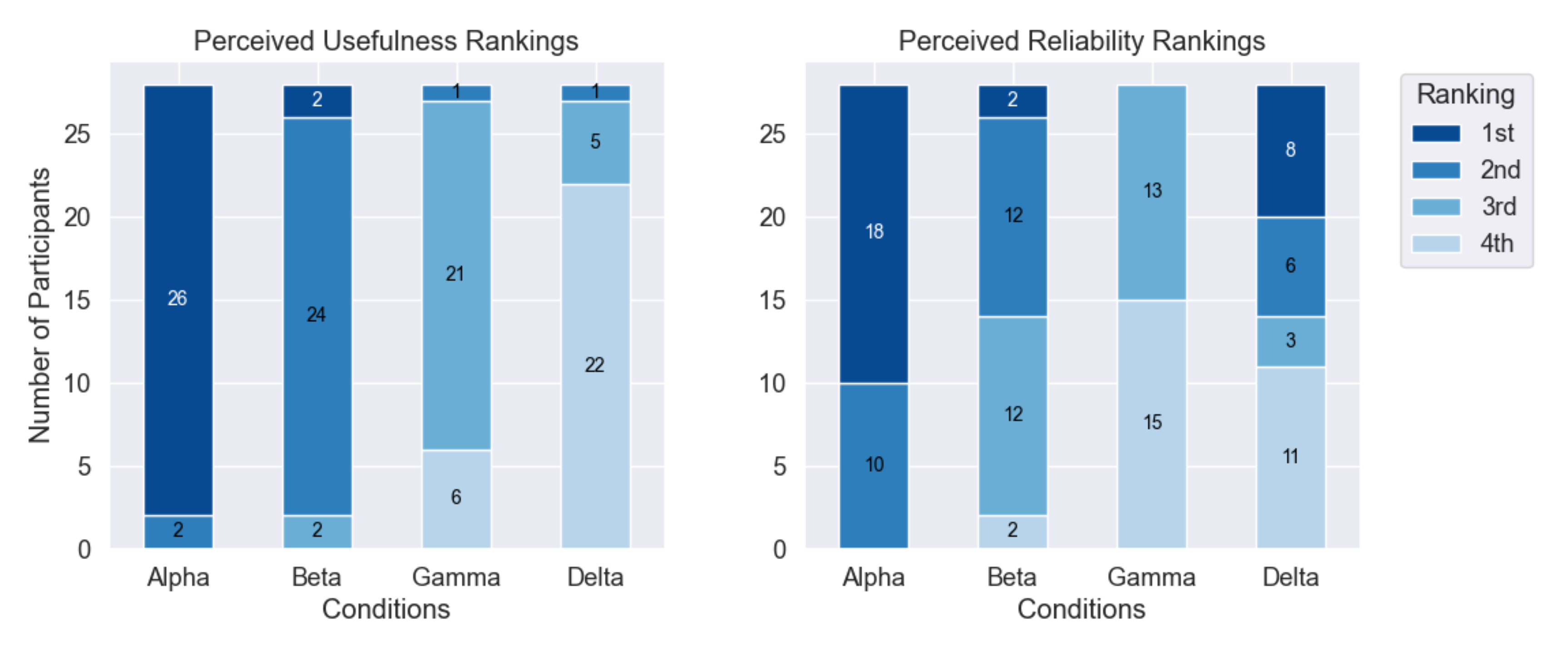}
    \vspace{-10pt}
    \caption{Perceived usefulness and reliability ranking: 1st represents the most useful/reliable, while 4th represents the least useful/reliable. The number on each bar indicates the number of participants who gave this ranking.
    }
    \Description{}
    \label{fig:figure5}
\end{figure*}

We present the results of post-session questionnaire in Figure~\ref{fig:figure5}.

\section{Details of Data Source and Data Processing}\label{sec:dataset_detail}
\subsection{Details of Data Source}
The Meta Kaggle and Meta Kaggle Code datasets span from 2015 to 2024, including a total of 4.83 million code files~\cite{kaggleMetaKaggle,metakagglecode}. 
Notably, 4.36 million of these files are Jupyter notebooks, which form the focus of our study. 
The datasets cover 5,688 competitions and include contributions from approximately 1.22 million users~\cite{kaggleMetaKaggle}.
In addition to the code files, the dataset also contains rich social features, such as count of views, votes, comments, and other forms of engagement, providing valuable insights into community interactions.
\subsection{Competition Details}\label{sec:competition}
Details about extracted competitions are listed in Table~\ref{tab:tab10}.
\begin{table*}[h]
    \centering
    \scriptsize
    \caption{Selected Competitions}
    \begin{tabular}{p{4cm}p{4cm}p{3cm}p{1cm}}
     \toprule
      \textbf{Competition} & \textbf{Description} & \textbf{Domain} & \textbf{Python Files} \\
      \midrule
       Quora Insincere Questions Classification  
       & Detect toxic content to improve online conversations 
       & Natural Language Processing & 9,639\\
       Ubiquant Market Prediction 
       & Make predictions against future market data 
       & Business Data Analysis & 4,529\\
       Cassava Leaf Disease Classification 
       & Identify the type of disease present on a Cassava Leaf image 
       & Computer Vision & 12,911\\
       Mechanisms of Action (MoA) Prediction 
       & Can you improve the algorithm that classifies drugs based on their biological activity?
       & Medical Data Analysis & 9,866\\
     \bottomrule
    \end{tabular}

    \label{tab:tab10}
\end{table*}
\subsection{Details of Data Processing} \label{sec:dataset_process}
In summary, we processed 37,895 code files and obtained 36,945 valid Python-written Jupyter Notebook files, containing 849,900 code cells and 269,855 markdown cells. 
For each valid file, we collect the following meta information: public post URL, title, vote count, view count, comment count, notebook submission date, author name, and the notebook author's public profile avatar.
Our data collection adheres strictly to Kaggle's policies, ensuring that all utilized information is open source and publicly accessible.

\section{\ac{RAG} Model}\label{sec:RAG-details}

\subsection{The Retriever Module} 
\subsubsection{Technical Details}
We begin by constructing comprehensive Kaggle post databases, creating a separate database for each Kaggle competition. 
Each notebook submission is divided into smaller chunks, consisting of a continuous group of markdown cells followed by corresponding code cells, up until the next markdown cell. This approach balances prompt length and retrieval efficiency—chunking the entire notebook risks exceeding the input length limits of LLMs and increasing generation time, while using individual code cells would miss the connection between markdown explanations and code. By chunking markdown and code together, users can retrieve relevant information using either text explanations or code.
On average, each chunk in our processed dataset contains 1.2 markdown cells and 3.6 code cells.

To encode these chunks into vector embeddings, we use OpenAI's \texttt{text-embedding-ada-002} ~\cite{openai_new_2024} model. 
The embedding process for a given chunk \(c_i\) is formally defined as\footnote{The exact implementation of the text-embedding-ada-002 model may vary, and components like pooling or linear projection may differ or be omitted.}:

\[
\mathbf{e}_{c_i} = \texttt{LinearProjection}\left(\texttt{Pool}\left(\texttt{TransformerLayers}\left(\texttt{TokenEmbedding}\left(\texttt{Tokenize}(c_i)\right)\right)\right)\right)
\]

where:
- \(\texttt{Tokenize}(c_i)\) converts the chunk \(c_i\) into a sequence of tokens,
- \(\texttt{TokenEmbedding}(\cdot)\) maps each token to an initial dense vector representation,
- \(\texttt{TransformerLayers}(\cdot)\) applies multiple transformer layers to capture contextual information across tokens,
- \(\texttt{Pool}(\cdot)\) aggregates the sequence of token embeddings into a single vector using pooling,
- \(\texttt{LinearProjection}(\cdot)\) maps the pooled vector to the final embedding space.

The resulting embedding \(\mathbf{e}_{c_i}\) captures the semantic content of the entire chunk, integrating information from both markdown text and associated code blocks. 
We use the same embedding for both markdown and code to ensure uniform retrieval, enabling the model to effectively retrieve relevant information regardless of the input text type.
These embeddings are then stored in a vector database, along with their corresponding chunk indices. 
For this study, we utilize ChromaDB~\cite{chromadb24} to manage and query these embeddings efficiently.

When a user inputs a query $q$, we calculate the \ac{MMR} score using the vector representation of the query \(\mathbf{q}\) and the embeddings of each chunk \(\mathbf{e}_{c_i}\) to identify the top 10 most relevant chunks. This process can be formally writen as:

\[
\text{MMR}(c_i, \mathbf{q}) = \arg\max_{c_i \in C} \left[\text{Relevance}(\mathbf{q}, \mathbf{e}_{c_i}) - \lambda \max_{c_j \in S} \text{Sim}(\mathbf{e}_{c_i}, \mathbf{e}_{c_j})\right]
\]

where \(C\) is the set of all chunks, \(S\) is the set of already selected chunks, \(\text{Relevance}(\mathbf{q}, \mathbf{e}_{c_i})\) represents the similarity between the query vector \(\mathbf{q}\) and chunk embedding \(\mathbf{e}_{c_i}\), and \(\lambda\) is a parameter controlling the diversity of the selected chunks. 
The choice of the top 10 is consistent with established practices in information retrieval~\cite{jones1999phrasier}.

These top 10 chunks are then ranked according to the user's preference using one of the following methods:

\begin{enumerate}
    \item \textbf{Relevance (MMR Score)}: The chunks are ranked by their \(\text{MMR}(c_i, \mathbf{q})\) scores, which can be expressed as:

    \[
    \text{Rank}_{\text{MMR}}(c_i) = \text{Sort}\left(\text{MMR}(c_i, \mathbf{q}), \text{descending}\right)
    \]

    \item \textbf{View Count}: The chunks are ranked according to the view count \(V_{c_i}\) of the corresponding post:

    \[
    \text{Rank}_{\text{View}}(c_i) = \text{Sort}(V_{c_i}, \text{descending})
    \]

    \item \textbf{Vote Count}: The chunks are ranked according to the vote count \(R_{c_i}\) of the corresponding post:

    \[
    \text{Rank}_{\text{Vote}}(c_i) = \text{Sort}(R_{c_i}, \text{descending})
    \]
\end{enumerate}

Based on the user's selection of the ranking method, top \(N\) chunks are selected from the sorted list:

\[
\{c_i\}_{\text{top } N} = \text{Rank}_{\text{Method}}(c_i)[1:N]
\]

where \(\text{Rank}_{\text{Method}}(c_i)\) denotes the list of chunks sorted by the selected ranking method (relevance, view count, or vote count). These top \(N\) chunks are then presented to the user along with links to their respective Kaggle notebook posts.

\subsection{The Generator Module} 
The retrieved posts, along with the user's query, are formulated into a comprehensive prompt \(P\):

\[
P = \text{Format}(\mathbf{q}, \{c_i\}_{\text{top } N})
\]

where \(\mathbf{q}\) is the user's query, and \(\{c_i\}_{\text{top } N}\) represents the top \(N\) retrieved chunks selected in the retriever module. The \(\text{Format}(\cdot)\) function constructs a structured prompt by integrating the user's query \(\mathbf{q}\) with the content of the top \(N\) chunks.

The prompt \(P\) is then fed into the GPT-4o model:

\[
\mathbf{r} = \text{GPT-4o}(P)
\]

where \(\mathbf{r}\) is the generated response. 
The GPT-4o model synthesizes the information from the prompt \(P\) to produce a coherent and tailored response \(\mathbf{r}\) that addresses the user's query. 

We use the following prompt for generating response:

\noindent\texttt{
Task: You are an expert in data science. Your goal is to assist a user in dealing with a task related to a Kaggle competition: \{title of competition\}. Whenever possible, provide answers in the form of code snippets that directly address the user's needs.
}

\noindent\texttt{Information Provided:
\begin{enumerate}
    \item Competition Description: \{One sentence description of the competition\}
    \item Context: \{context\}
\end{enumerate}
}

\noindent\texttt{
*Note:* The context includes other people's code, which contains information necessary for answering the user's question. Please rely solely on the provided context to craft your response. Assume all questions pertain specifically to this competition. If the context is empty, state \say{There is no relevant information in previous notebooks} and then proceed to answer the question based on your expertise.
}

\noindent\texttt{User Query: \{question\}}

\noindent\texttt{Expected Output: Please provide your response as a string, including code snippets where applicable.}



\begin{table*}[h]
\centering
\scriptsize
\caption{User behavior statistics.}
\renewcommand{\arraystretch}{1.8} 
\begin{tabular}{>{\raggedright\arraybackslash}m{3.5cm} p{1cm} p{0.3cm} p{1cm} p{0.5cm} p{5.5cm}}
\toprule
\textbf{Attribute} & \textbf{Cond.} & \textbf{N} & \textbf{Mean} & \textbf{SD} \\
\midrule
\multirow{3}{4cm}{Input Prompts} 
& Alpha & 28 & 5.25 & 2.84 & \multirow{3}{*}{\includegraphics[width=5.5cm,height=1.5cm]{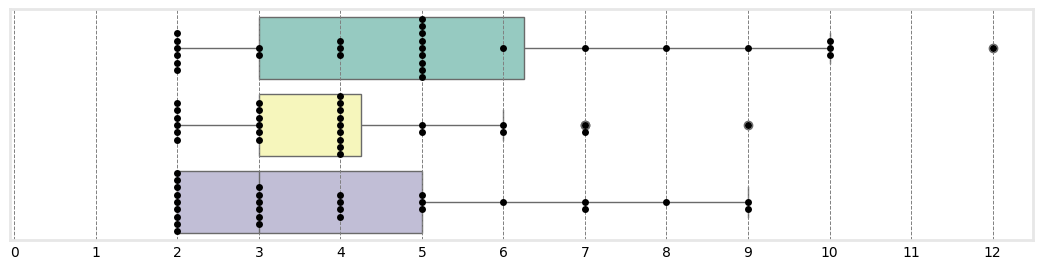}}\\
& Beta & 28 & 3.96 & 1.75 \\
& Gamma & 28 & 4.04 & 2.22 \\
\midrule
\multirow{1}{4cm}{Times Clicking Preview} 
& Alpha & 28 & 2.75 & 2.55 & {\includegraphics[width=5.5cm,height=0.5cm]{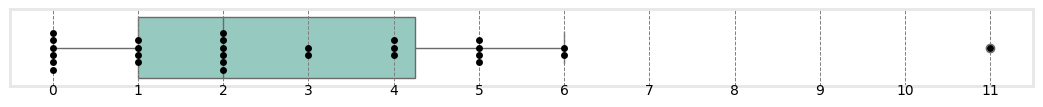}}\\
\midrule
\multirow{1}{4cm}{Times Using Advanced Search Panel} 
& Alpha & 28 & 1.71 & 1.82 & {\includegraphics[width=5.5cm,height=0.5cm]{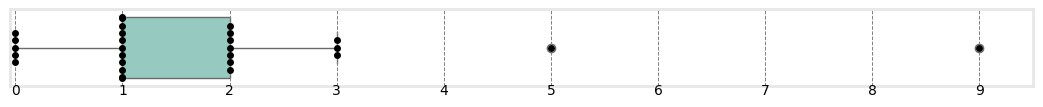}} \\
\bottomrule
\end{tabular}

\label{tab:user_behavior}
\end{table*}
\section{Detailed Analysis of User Interactions}\label{sec:user_behavior}
We present user behavior statistics across different conditions in Table~\ref{tab:user_behavior}. 


\received{May 2025}
\received[revised]{November 2025}
\received[accepted]{December 2025}
\end{document}